\documentclass[lettersize,journal]{IEEEtran}
\usepackage{amsmath,amsfonts,amssymb}
\usepackage{amsthm}
\usepackage{mathtools}
\usepackage{array}
\usepackage{textcomp}
\usepackage{stfloats}
\usepackage{url}
\usepackage{graphicx}
\usepackage{cite}
\graphicspath{{figures/}}
\newtheorem{theorem}{Theorem}
\newtheorem{proposition}{Proposition}
\newtheorem{drule}{Design Rule}
\newcommand{\dd}{\mathrm d}
\DeclareMathOperator{\asinh}{asinh}
\DeclareMathOperator{\acosh}{arcosh}
\DeclareMathOperator{\atanh}{atanh}
\DeclareMathOperator{\sech}{sech}
\DeclareMathOperator{\rank}{rank}
\renewcommand{\Re}{\operatorname{Re}}

\newcommand{\IL}{\mathrm{IL}}
\newcommand{\See}{S_{\mathrm{ee}}}
\newcommand{\Set}{S_{\mathrm{et}}}
\newcommand{\Ste}{S_{\mathrm{te}}}
\newcommand{\Stt}{S_{\mathrm{tt}}}
\newcommand{\SN}[1]{Supplementary Note~S#1}

\begin{document}
\title{A Design Rule for Passive Reconfigurable\\ Microwave Devices}
\author{Ali~Nadarloo, Mohammad~Javad~Hajiahmadi, and Mohammad~Soleimani\thanks{The authors are with the School of Electrical Engineering, Iran University of Science and Technology, Tehran 1684613114, Iran.}\thanks{Corresponding author: Mohammad Javad Hajiahmadi (e-mail: hajiahmadi@iust.ac.ir).}}

\maketitle

\begin{abstract}
Every reconfigurable microwave device places a tuning element---a varactor, a diode, transistor, microelectromechanical or phase-change switch, a ferroelectric or liquid-crystal load---inside a fixed passive network. The network can make a small change of the element produce a large change of the response, and design practice often credits it with the device's tuning capability. This paper shows that a fixed, linear, passive network cannot add to that capability. Each tuning element carries a tuning budget, a dimensionless length computed from its own impedance trajectory, and a fixed passive network can transmit that budget or lose part of it, never enlarge it. The unit of the budget is the hyperbolic distance of the Smith chart: the distance from a match is the logarithm of the voltage standing-wave ratio, and the distance between two states sets their mutual mismatch and is the logarithm of their commutation quality factor. From one inequality we derive closed-form, ready-to-use limits for analog and digital phase shifters, true-time-delay lines, switches, attenuators, vector modulators, impedance tuners, tunable filters, reconfigurable antennas, metasurface cells and multiport networks; the two-state and attenuation limits are attained by explicit circuits, and the phase limit by a constant-loss trajectory. We introduce the $\Lambda$-chart, a Smith-chart companion on which budgets and requirements are read graphically, and apply it to four published microwave case studies.
\end{abstract}

\begin{IEEEkeywords}
Commutation quality factor, hyperbolic geometry, impedance tuners, phase shifters, Smith chart, switches.
\end{IEEEkeywords}

\section{Introduction}
\IEEEPARstart{A}{sk} a microwave designer how to get more phase range from a varactor and the answer is usually a better circuit around it: a resonant load, a matching section, a coupled-line transformer, a reflection-type topology. Near resonance a fraction of a picofarad swings the reflection phase through most of a circle, and the sensitivity of the response to the element becomes very large. It is tempting to conclude that a sufficiently clever circuit can turn any tuning element into any reconfigurable function.

Yet a low-$Q$ varactor does not become a low-loss $360^\circ$ phase shifter however well it is matched; a reconfigurable-intelligent-surface (RIS) cell with one varactor shows a deep amplitude dip exactly where its phase changes fastest; and at millimeter waves the topology that worked at a few gigahertz runs out of tuning. Classical theory explains part of this. Hines showed that the loss of a diode phase shifter has a floor set by the diode's cutoff frequency \cite{Hines1964}, Kawakami found a figure of merit of a two-state one-port that no lossless reciprocal network can change \cite{Kawakami1965}, and Vendik \emph{et al.} turned it into the commutation quality factor (CQF) now used to rank switching elements \cite{Vendik2000}. These results concern two states. Modern reconfigurable hardware is continuous, multiport, or both: analog phase shifters that sweep a full turn, impedance tuners that must cover part of the Smith chart, metasurface cells, pattern-reconfigurable antennas and programmable networks with thousands of states.

This paper gives a single rule that covers these cases. A tuning element carries a \emph{tuning budget} $\Lambda$, a dimensionless length computed from its datasheet before any circuit is drawn. A specification carries a \emph{tuning requirement}, computed from the response it asks for. A fixed passive network delivers at most the budget it receives. When the requirement exceeds the budget, no matching section, resonance, coupler, cascade or feed network meets the specification; the designer must change the element, the frequency or the specification.

The mathematics behind the rule is classical. Passive scattering matrices are contractions \cite{Youla1959,Kurokawa1965}, and contractive analytic maps cannot stretch the invariant geometry of the unit ball \cite{Kobayashi1967,Harris1979}; Helton showed long ago that this geometry belongs in circuit theory \cite{Helton1982}, Allen and Healy used it for impedance matching \cite{Allen2004}, and the hyperbolic structure of the Smith chart is well known \cite{Smith1939,Ohira2020,Muller2011}. The contributions are on the engineering side:
\begin{enumerate}
\item an access--return factorization that shows where a passive network loses tuning motion, and a physical reading of the budget as \emph{contrast}: the reflection one state presents when the circuit is matched to the other (Sections~\ref{sec:why}--\ref{sec:meaning});
\item closed-form budgets of practical tuning elements, including distributed and material tuners (Section~\ref{sec:budgets});
\item ready-to-use design rules for the main classes of reconfigurable microwave hardware, with circuits that attain them, and an exact geometric law for the amplitude--phase coupling of single-element metasurface cells (Section~\ref{sec:rules});
\item the $\Lambda$-chart, a graphical tool built on the Smith chart for reading budgets and requirements (Section~\ref{sec:chart});
\item a map of how the classical limits fit into the rule (Section~\ref{sec:prior}) and four published microwave case studies (Section~\ref{sec:cases}).
\end{enumerate}
The Supplementary Information gives the derivations step by step (Supplementary Notes S1--S16), a user's manual for the $\Lambda$-chart (\SN{17}), worked examples from published devices (\SN{18}), the optical, acoustic and quantum counterparts (\SN{19}), the scope conditions (\SN{20}) and numerical checks of the main results (\SN{21}).

\section{Why a Passive Network Cannot Create Tuning}\label{sec:why}

\subsection{The Partition}
We use power waves \cite{Kurokawa1965}, so that $|a|^2$ and $|b|^2$ are incident and outgoing powers and a passive multiport is a contraction, $S^\dagger S\preceq I$ \cite{Youla1959}. Every reconfigurable device is split into two parts (Fig.~\ref{fig:concept}b):
\begin{itemize}
\item the \emph{tuner} $\Gamma(q)$, the scattering matrix at its own ports of everything that changes with the control $q$ (bias, actuation, temperature, light);
\item the \emph{network} $S$, everything that stays fixed: lines, couplers, hybrids, resonators, matching sections, package, bias network and radiating structure.
\end{itemize}
With external ports (subscript e) and tuner ports (subscript t),
\begin{equation}
\begin{bmatrix}b_{\mathrm e}\\ b_{\mathrm t}\end{bmatrix}
=\begin{bmatrix}\See&\Set\\ \Ste&\Stt\end{bmatrix}
\begin{bmatrix}a_{\mathrm e}\\ a_{\mathrm t}\end{bmatrix},\qquad a_{\mathrm t}=\Gamma b_{\mathrm t},
\end{equation}
and the response of the device is the terminated-network formula
\begin{equation}
F(\Gamma)=\See+\Set\Gamma(I-\Stt\Gamma)^{-1}\Ste .
\label{eq:F}
\end{equation}
For one element, $\Gamma$ is its reflection coefficient and \eqref{eq:F} is the familiar input reflection or transmission of a terminated two-port \cite{Pozar2012}. For $m$ independent lumped elements, $\Gamma$ is diagonal.

Three conventions are used throughout. Anything that changes with $q$, including bias-dependent parasitics, belongs to the tuner. A fixed lossless element, such as package inductance, may sit on either side of the boundary without changing anything. A fixed lossy element next to the tuner, such as a contact resistance, may be absorbed into the tuner; this lowers the budget and gives a tighter limit (\SN{6}).

\subsection{Two States: Contrast}
Start with two settings of one element, reflection coefficients $\gamma_1$ and $\gamma_2$. Suppose a lossless network is designed to match the element in state~1. In state~2 the matched port then sees the reflection
\begin{equation}
r_{\mathrm c}=\left|\frac{\gamma_2-\gamma_1}{1-\gamma_1^*\gamma_2}\right| ,
\label{eq:contrast}
\end{equation}
because matching state~1 is a disk automorphism that carries $\gamma_1$ to the origin, and every lossless two-port acts on reflection coefficients as such an automorphism (\SN{2}). We call $r_{\mathrm c}$ the \emph{contrast} of the two states (Fig.~\ref{fig:concept}c). It does not depend on which lossless network is used: rotating with a line, transforming with a transformer, or adding reactances moves both states but leaves $r_{\mathrm c}$ unchanged. This is Kawakami's invariance \cite{Kawakami1965}.

A lossy network does worse. The Schwarz--Pick lemma states that an analytic map of the unit disk into itself cannot increase the quantity \eqref{eq:contrast} between any two points \cite{Ahlfors1973}. Every passive terminated response \eqref{eq:F} is such a map of $\gamma$. Therefore:
\begin{quote}\emph{No passive network can make two states of a tuning element look more different than its contrast allows.}\end{quote}

\subsection{From Two States to a Trajectory}
For nearby states, $r_{\mathrm c}\to|\dd\gamma|/(1-|\gamma|^2)$. Accumulating twice this contrast along the trajectory of the element defines its length
\begin{equation}
\dd\Lambda=\frac{2|\dd\gamma|}{1-|\gamma|^2}=\frac{|\dd Z|}{\Re Z},
\label{eq:metric1}
\end{equation}
which is the hyperbolic (Poincar\'e) line element of the reflection disk and of the right half impedance plane \cite{Beardon1983}; the reference impedance cancels (\SN{5}). The distance $d$ between two states is the length of the shortest trajectory joining them, and $r_{\mathrm c}=\tanh(d/2)$. A reactance change is therefore worth much when the element is nearly lossless and little when it is lossy, and a trajectory that touches the lossless boundary has infinite length. The \emph{tuning budget} of an element is $\Lambda$ evaluated along its trajectory over the allowed control range.

For a multiport tuner, defect matrices $D_X=(I-X^\dagger X)^{1/2}$ and $D_{X^\dagger}=(I-XX^\dagger)^{1/2}$ replace $1-|\gamma|^2$:
\begin{equation}
\dd\Lambda_X=2\bigl\|D_{X^\dagger}^{-1}\,\dd X\,D_X^{-1}\bigr\|_2=\bigl\|R^{-1/2}\,\dd Z\,R^{-1/2}\bigr\|_2 ,
\label{eq:metric}
\end{equation}
with $R=\tfrac12(Z+Z^\dagger)$ the resistance matrix and $\|\cdot\|_2$ the largest singular value. This is the Carath\'eodory--Kobayashi metric of the matrix ball \cite{Harris1979,Helton1982}.

\subsection{The Rule}
\begin{theorem}[Passive reconfiguration theorem]\label{thm:main}
Let $S$ be passive and let $\Gamma$ and $F(\Gamma)$ be strictly passive. Then
\begin{equation}
\mathcal V_F=P_\Gamma\,\mathcal V_\Gamma\,Q_\Gamma,\qquad \|P_\Gamma\|_2\le1,\quad\|Q_\Gamma\|_2\le1,
\label{eq:factor}
\end{equation}
where $\mathcal V_X=D_{X^\dagger}^{-1}\dd X\,D_X^{-1}$,
$Q_\Gamma=D_\Gamma(I-\Stt\Gamma)^{-1}\Ste D_F^{-1}$ and
$P_\Gamma=D_{F^\dagger}^{-1}\Set(I-\Gamma\Stt)^{-1}D_{\Gamma^\dagger}$. Consequently, for every trajectory and every pair of settings $X,Y$,
\begin{align}
\dd\Lambda_F&\le\dd\Lambda_\Gamma,\qquad \Lambda_F\le\Lambda_\Gamma,\label{eq:law}\\
d\bigl(F(X),F(Y)\bigr)&\le d(X,Y),\label{eq:dist}\\
\rank\bigl[F(Y)-F(X)\bigr]&\le\rank(Y-X).\label{eq:rank}
\end{align}
The same holds for any selected part of the response: one coefficient $v^\dagger Fu$, the output vector $Fu$ for a fixed excitation $u$, or any sub-block.
\end{theorem}

The proof (\SN{4}) rests on two energy balances,
\begin{equation}
\begin{aligned}
D_F^2&=\Ste^\dagger R_\Gamma^\dagger D_\Gamma^2R_\Gamma\Ste+T_R^\dagger(I-S^\dagger S)T_R,\\
D_{F^\dagger}^2&=\Set L_\Gamma D_{\Gamma^\dagger}^2L_\Gamma^\dagger\Set^\dagger+T_L^\dagger(I-SS^\dagger)T_L,
\end{aligned}
\label{eq:balance}
\end{equation}
with $R_\Gamma=(I-\Stt\Gamma)^{-1}$, $L_\Gamma=(I-\Gamma\Stt)^{-1}$ and $T_R,T_L$ the internal-wave maps. In words: the power lost by the device equals the power lost in the tuner, seen through the network, \emph{plus} the power dissipated in the network on the excited waves. The second term is never negative, because the network is passive. This is where the limit comes from. Dividing the motion of the tuner by its own defect and the motion of the response by its own defect gives the factors $Q_\Gamma$ (how well the incident waves reach the tuner) and $P_\Gamma$ (how much of the tuner's motion returns to the observed ports), and both are at most one. Motion is lost in two ways: by dissipation on the waves that carry it, and when the network does not bring the incident wave to the changing part of the tuner or does not return that part to the observed port; a tuner port left uncoupled can move without moving the response at all. A lossless, coupled two-port loses nothing, and nothing makes either factor larger than one.

\begin{figure*}[!t]
\centering
\includegraphics[width=\textwidth]{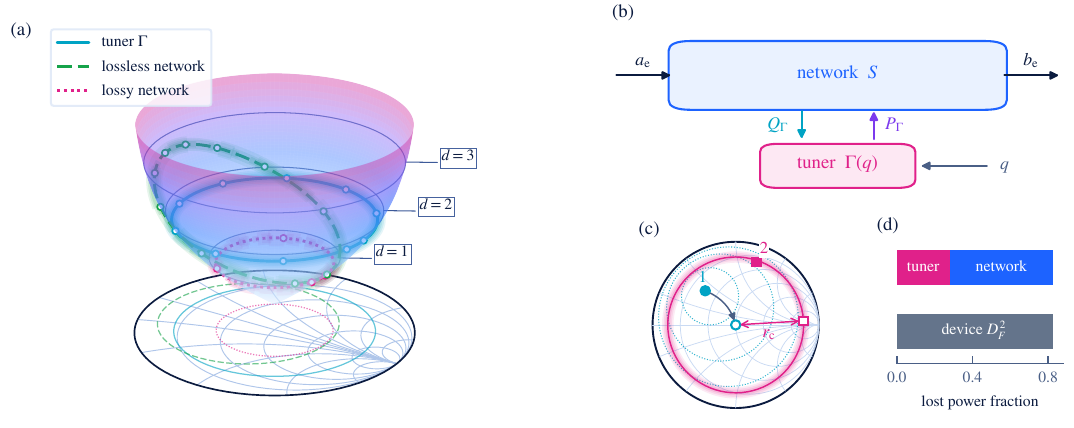}
\caption{The tuner, the network and the budget. (a) Budget surface: the Smith chart (floor) lifted to the height $d=\ln\mathrm{VSWR}$, the distance from a match; the rings $d=1,2,3$ correspond to VSWR $2.7$, $7.4$ and $20$. Dots mark every $\Delta\Lambda=2$ of budget length. The tuner trajectory $\Gamma=0.72e^{j\theta}$ has $\Lambda=18.79$; after a lossless network it is moved and reshaped and keeps $\Lambda=18.79$; after the lossy network $F=0.58\Gamma$ it keeps only $\Lambda=6.36$. The surface steepens toward the rim, so a given budget gives less Euclidean motion near the lossless boundary. (b) Partition of a device into the fixed passive network $S$ (lines, couplers, resonators, matching, package, bias network, radiator) and the tuner $\Gamma(q)$ driven by the control $q$; $Q_\Gamma$ and $P_\Gamma$ are the access and return factors of \eqref{eq:factor}. (c) Contrast: when a lossless network matches state~1 (moved to the center), state~2 appears at radius $r_{\mathrm c}=\tanh(d/2)$; dotted circles are $d=1,2,3$ about state~1, and the solid ring is $|\gamma|=r_{\mathrm c}$. (d) Energy balance \eqref{eq:balance} for the lossy example of (a), realized with $S_{\mathrm{et}}=S_{\mathrm{te}}=\sqrt{0.58}$ and $S_{\mathrm{ee}}=S_{\mathrm{tt}}=0$; the split depends on this choice and is the same at all states of the trajectory $|\Gamma|=0.72$: the device loss $D_F^2=0.83$ equals the tuner loss seen through the network ($0.28$) plus the network loss ($0.55$), which is why $\|P_\Gamma\|_2,\|Q_\Gamma\|_2\le1$.}
\label{fig:concept}
\end{figure*}

The factorization separates two quantities that design practice often merges. The lossless transformer embedding $f(\gamma)=(\gamma-a)/(1-a\gamma)$ magnifies the ordinary derivative at $\gamma=a$ by $1/(1-a^2)$, without bound as $a\to1$, yet its intrinsic gain $\dd\Lambda_F/\dd\Lambda_\Gamma$ is exactly one everywhere (Fig.~\ref{fig:concept}a, lossless-network curve). The attenuating embedding $F=r\gamma$, $0<r<1$, has derivative $r$ and shortens every trajectory (Fig.~\ref{fig:concept}a, lossy-network curve); no later lossless stage restores what it removed. \emph{Sensitivity measures how fast a response moves; the budget measures how far it can move.}

\section{What the Budget Means in Microwave Terms}\label{sec:meaning}
\begin{proposition}\label{prop:meaning}
For reflection coefficients and impedances with positive resistance (\SN{3}):
\begin{enumerate}
\item \emph{VSWR.} $d(0,\gamma)=2\atanh|\gamma|=\ln\mathrm{VSWR}$.
\item \emph{Mismatch.} A source $\Gamma_{\mathrm s}$ delivers to a load $\Gamma_{\mathrm L}$ the fraction
\begin{equation}
\eta_{\mathrm m}=\frac{(1-|\Gamma_{\mathrm s}|^2)(1-|\Gamma_{\mathrm L}|^2)}{|1-\Gamma_{\mathrm s}\Gamma_{\mathrm L}|^2}=\sech^2\frac{d(\Gamma_{\mathrm s}^*,\Gamma_{\mathrm L})}{2}
\label{eq:mismatch}
\end{equation}
of its available power.
\item \emph{CQF.} Two impedances $Z_1,Z_2$ have Kawakami's invariant $r_{\mathrm c}=|(Z_1-Z_2)/(Z_1+Z_2^*)|$ \cite{Kawakami1965} and the commutation quality factor $K$ of Vendik \emph{et al.} \cite{Vendik2000}, with
\begin{equation}
\begin{aligned}
K&=\frac{1+r_{\mathrm c}}{1-r_{\mathrm c}}=e^{d},\\
\kappa^2&\equiv\frac{|Z_1-Z_2|^2}{\Re Z_1\,\Re Z_2}=K+\frac1K-2=4\sinh^2\frac d2.
\end{aligned}
\label{eq:CQF}
\end{equation}
The switching ratio $\kappa=\sqrt K-1/\sqrt K$ reduces to the cutoff ratio $f_c/f$ for a diode or switch in the reactive-dominated regime \cite{Hines1964,White1974}; the general two-state expression is given in Table~\ref{tab:budgets}. $K$ is the CQF as defined in \cite{Vendik2000}; the ferroelectric form $(n-1)^2/(n\tan\delta_1\tan\delta_2)$ approximates $\kappa^2$, which is close to $K$ only at high contrast.
\item \emph{Two outputs.} For any two scalar responses $t_1,t_2$,
\begin{equation}
\sinh^2\frac{d(t_1,t_2)}{2}=\frac{|t_1-t_2|^2}{(1-|t_1|^2)(1-|t_2|^2)}.
\label{eq:twostate}
\end{equation}
\end{enumerate}
\end{proposition}
By \eqref{eq:mismatch}, the mismatch loss is $20\log_{10}\cosh(d/2)$\,dB. One unit of length is the distance from a matched port to VSWR $e\approx2.72$, and two states one unit apart mismatch each other by $1.04$\,dB. The classical invariance of the CQF under lossless reciprocal embedding \cite{Kawakami1965,Vendik2000} is the lossless two-point case of \eqref{eq:dist}. Theorem~\ref{thm:main} extends it to trajectories, lossy and nonreciprocal networks, multiports and sets of states. The extension matters: a full $360^\circ$ phase cycle begins and ends at the same response, so every two-state measure misses it, yet it consumes a definite length.

\section{Budgets of Tuning Elements}\label{sec:budgets}
Integrating \eqref{eq:metric1} gives the budgets in Table~\ref{tab:budgets} (derivations in \SN{6}). We write $n=C_{\max}/C_{\min}$ and $\omega=2\pi f$.

\begin{table*}[!t]
\centering
\caption{Tuning Budgets of Microwave Tuning Elements at Frequency $f$}
\label{tab:budgets}
\renewcommand{\arraystretch}{1.3}
\small
\begin{tabular}{@{}>{\raggedright\arraybackslash}p{0.44\textwidth}>{\raggedright\arraybackslash}p{0.22\textwidth}>{\raggedright\arraybackslash}p{0.28\textwidth}@{}}
\hline
Element and model & Path budget $\Lambda$ & $\kappa^2=K+1/K-2$ of end states\\
\hline
Varactor or MEMS capacitor, $Z=R_s-j/(\omega C)$; $Q_{\max}=1/(\omega R_sC_{\min})$ &
$Q_{\max}(1-1/n)$\newline $=Q_{\max}-Q_{\min}$ & $\Lambda^2$\\
Constant-$Q$ capacitor or inductor (ferroelectric, liquid crystal, ferrite), constant loss tangent $1/Q$ &
$\sqrt{1+Q^2}\,\ln n$ & $(1+Q^2)(n-1)^2/n$\\
Measured or modeled $Z(V)$ &
$\int|Z'(V)|/\Re Z\,\dd V$ & $|Z_1-Z_2|^2/(R_1R_2)$\\
Switch: $Z_{\mathrm{on}}=R_{\mathrm{on}}$, $Z_{\mathrm{off}}=R_{\mathrm{off}}-j/(\omega C_{\mathrm{off}})$; $f_c=1/(2\pi C_{\mathrm{off}}\sqrt{R_{\mathrm{on}}R_{\mathrm{off}}})$ &
--- & $(f_c/f)^2+\frac{(R_{\mathrm{off}}-R_{\mathrm{on}})^2}{R_{\mathrm{on}}R_{\mathrm{off}}}$\\
Multiport tuner $Z(q)$ & $\int\|R^{-1/2}Z'R^{-1/2}\|_2\dd q$ & from $d$, \SN{4}\\
\hline
\end{tabular}
\end{table*}

\emph{Stacking neither costs nor adds budget.} $k$ identical elements in series or parallel scale $Z$ by $k$ or $1/k$, and $|\dd Z|/\Re Z$ is unchanged.

\emph{Jointly driven identical elements count once.} They form $\Gamma=\gamma I_k$, whose length is that of one element. Their extra value is structural (Rules~\ref{rule:count} and~\ref{rule:multi}).

\emph{Independent elements combine by maximum.} For $\Gamma=\bigoplus_k\Gamma_k$, the line element is $\dd\Lambda=\max_k\dd\Lambda_k$ and the distance between two settings is $d=\max_kd_k$. When the elements move together, the path length therefore lies between the largest single length and the sum of the lengths.

\emph{Jointly driven distributed tuners have the budget of one cell.} A loaded line or a tunable dielectric is a set of identical, jointly driven cells whose impedances are scaled copies of each other; since the budget is scale invariant, the whole structure has the budget of one cell. This holds while the material acts through its local capacitance along a fixed field direction; a material whose principal axes rotate couples polarizations and is treated with the matrix budget \eqref{eq:metric} (\SN{6}). For a tunable dielectric with permittivity range $\varepsilon_{\mathrm r,\min}$--$\varepsilon_{\mathrm r,\max}$ and bias-dependent loss tangent $\tan\delta(\varepsilon_{\mathrm r})=1/Q(\varepsilon_{\mathrm r})$,
\begin{equation}
\begin{aligned}
\Lambda&=\int_{\varepsilon_{\mathrm r,\min}}^{\varepsilon_{\mathrm r,\max}}\sqrt{Q^2+\Bigl(1+\frac{\dd\ln\tan\delta}{\dd\ln\varepsilon_{\mathrm r}}\Bigr)^2}\,\frac{\dd\varepsilon_{\mathrm r}}{\varepsilon_{\mathrm r}},\\
\Lambda_{Q\,\mathrm{const}}&=\sqrt{1+Q^2}\,\ln\frac1{1-\tau}\ \approx\ \frac{\tau}{\tan\delta}\quad(\tau\ll1,\ Q\gg1),
\end{aligned}
\label{eq:material}
\end{equation}
with tunability $\tau=1-\varepsilon_{\mathrm r,\min}/\varepsilon_{\mathrm r,\max}$. The derivative term accounts for the change of dissipation along the sweep. The last form is the material figure of merit used for liquid crystals \cite{Jakoby2020}; the CQF of ferroelectric capacitors \cite{Vendik2000,Tagantsev2003} is its two-state counterpart. Conductor loss is part of the fixed network and can only lower the delivered length.

\emph{Frequency scaling is explicit.} A varactor budget falls as $1/f$; a switch's $\kappa$ falls as $1/f$ while its off-state reactance dominates, after which the frequency-independent resistance term of Table~\ref{tab:budgets} takes over; a constant-$Q$ budget follows $Q(f)$. These expressions provide frequency-dependent feasibility tests; a finite frequency ceiling occurs when the available resource falls below the corresponding requirement.

The budget can be measured directly: two calibrated VNA measurements of the tuner give $d$ from \eqref{eq:contrast} or its matrix form, and summing the distances between the states of a bias sweep gives a lower estimate of $\Lambda$ (\SN{6}). A lower estimate cannot exclude a design: an exclusion test needs an upper bound, taken from a device model or from monotonicity between the measured points (\SN{18}, E3).

\section{Design Rules}\label{sec:rules}
Each rule compares a requirement, computed from the specification alone, with a budget from Section~\ref{sec:budgets}. Each rule gives a necessary condition under its stated element, response and control assumptions for a fixed, linear, time-invariant passive network operating in the small-signal regime (Section~\ref{sec:scope}); no particular embedding topology is assumed. Losses are in nepers, $\ell=-\ln|t|$, $\IL=8.686\,\ell$\,dB, and include power absorbed, reflected or routed to unobserved ports. Table~\ref{tab:summary} collects the rules for direct use.

\subsection{Analog Phase Shifters and True-Time Delay}
For a selected channel $t=e^{-\ell+j\phi}$, \eqref{eq:metric1} becomes $\dd\Lambda_t=\sqrt{\dd\ell^2+\dd\phi^2}/\sinh\ell$: each radian of phase costs $1/\sinh\ell$ of budget, little at high loss and much at low loss.

\begin{drule}[Phase range versus loss]\label{rule:phase}
Let the unwrapped phase of a phase shifter travel a total of $\Phi$ rad. This total variation equals the phase range when the phase is monotone and exceeds it otherwise. Then
\begin{equation}
\Lambda\ge\frac{\Phi}{\sinh\ell_{\max}},\quad
\bar\ell_\phi\ge\asinh\frac{\Phi}{\Lambda},\quad
\Phi_{\ell\le\ell_0}\le\Lambda\sinh\ell_0,
\label{eq:phase}
\end{equation}
where $\bar\ell_\phi$ is the phase-weighted mean loss and $\Phi_{\ell\le\ell_0}$ the phase accumulated while the loss stays below $\ell_0$ (\SN{7}). Equality holds for a constant-loss circular trajectory transferred without loss, e.g.\ $\Gamma=\rho e^{j\theta}$ behind a circulator, for which $\Lambda=2\pi/\sinh(-\ln\rho)$; the bound is therefore sharp, although a particular element trajectory may not reach it (Rule~\ref{rule:horo}).
\end{drule}

The familiar figure of merit is $\mathrm{FoM}=\Phi_r^\circ/\IL_{\max}=6.596\,\Phi_r/\ell_{\max}$, with $\Phi_r$ the relative phase shift in radians. Since $\Phi_r\le\Phi$,
\begin{equation}
\mathrm{FoM}\le6.596\,\Lambda\,\frac{\sinh\ell_{\max}}{\ell_{\max}}\approx6.6\,\Lambda\quad({}^\circ/\mathrm{dB}).
\label{eq:fom}
\end{equation}
\emph{No topology, and no cascade of stages, gives a phase shifter a figure of merit above $6.596\Lambda\sinh\ell_{\max}/\ell_{\max}$;} at low loss this is $6.6$ times the budget of its tuning element, and the factor $\sinh\ell_{\max}/\ell_{\max}$ is $1.005$ at $1.56$\,dB and $1.24$ at $10$\,dB. In the present notation, Vendik and Parnes give $6.6\kappa$ for a line with tunable phase velocity \cite{Vendik2008}, i.e.\ $6.6\Lambda$ for a varactor; their larger $8.35\kappa$ for digital shifters applies to independently switched bits, not to one control path, and follows from Rule~\ref{rule:two}. With a varactor,
\begin{equation}
\begin{aligned}
\IL_{\max}&\ge8.686\asinh\frac{\Phi}{Q_{\max}(1-1/n)}\ \mathrm{dB},\\
f&\le f_{cv}\Bigl(1-\frac1n\Bigr)\frac{\sinh\ell_{\max}}{\Phi},
\end{aligned}
\label{eq:ceiling}
\end{equation}
with $f_{cv}=1/(2\pi R_sC_{\min})$. The second inequality is a frequency ceiling for a given diode and specification. A constant-$Q$ capacitor with $Q=50$ and $n=2$ has $\Lambda=34.7$: a full turn then costs at least $1.57$\,dB peak loss, and at this peak loss the figure of merit is at most $230^\circ$/dB. For a passive phase shifter with matched output at ambient temperature the noise figure equals the insertion loss \cite{Pozar2012}, so \eqref{eq:phase} is also a noise-figure floor for receive arrays.

A true-time-delay line that must provide a delay change $\Delta\tau$ up to frequency $f_H$ needs phase travel $2\pi f_H\Delta\tau$ at $f_H$. With a varactor, the budget at $f_H$ gives the delay--bandwidth--loss limit
\begin{equation}
\Delta\tau\,f_H^2\le\frac{f_{cv}(1-1/n)\sinh\ell_{\max}}{2\pi}.
\label{eq:ttd}
\end{equation}
For $f_{cv}=350$\,GHz, $n=6$ and $1$\,dB peak loss, a delay change above $53.6$\,ps is impossible at $f_H=10$\,GHz in any passive topology.

\subsection{How Many Elements a Phase Range Needs}
\begin{drule}[Element count]\label{rule:count}
Let $N$ tuning elements have impedances that are linear-fractional functions of one common control (identical varactors, constant-$R_s$ or constant-$Q$ models), let the control sweep monotonically, and let the selected response coefficient stay nonzero. Then
\begin{equation}
\Phi<N\cdot360^\circ .
\label{eq:count}
\end{equation}
\end{drule}
The response is then a ratio of polynomials of degree at most $N$ in the control, and each linear factor turns by less than $180^\circ$ over a real interval (\SN{8}). A phase range beyond $360^\circ$ needs at least two elements, as found for one tunable component in \cite{Vendik2008}; in randomly generated lossless networks, one, two and three elements reached at most $351^\circ$, $512^\circ$ and $597^\circ$ (\SN{21}).

\subsection{The Horocycle Law of Single-Element Cells}
For a lossless reflecting metasurface or reflectarray cell with one tuning element and one external propagating channel, the fixed structure can be represented as a lossless two-port terminated by the element. Its reflection is therefore a disk automorphism of the element's reflection (\SN{2}). A constant-$R_s$ varactor moves on the line $\Re Z=R_s$, which in the reflection disk is a \emph{horocycle}, a circle tangent to the unit circle. Automorphisms map horocycles to horocycles and preserve arc length.

\begin{drule}[Horocycle law]\label{rule:horo}
Let a reflecting cell have one propagating channel (one Floquet mode in one polarization), a lossless fixed structure and one constant-$R_s$ varactor coupled to that channel. Its reflection traces a horocycle segment of hyperbolic length $\Lambda=Q_{\max}(1-1/n)$. A horocycle of Euclidean radius $s$ that encloses the origin ($\tfrac12<s<1$) reaches amplitude $1$ at its tangent point and $2s-1$ opposite to it. Every horocycle segment of length $\Lambda$ is realized by some lossless cell, and a segment symmetric about the point nearest the origin, $1-s+se^{j\vartheta}$ with $\vartheta\in[\vartheta_1,2\pi-\vartheta_1]$ (tangent point rotated to $\gamma=1$), has
\begin{equation}
\begin{aligned}
\Lambda&=\frac{2\cot(\vartheta_1/2)}{1-s},\qquad |t|_{\min}=2s-1,\\
\Phi&=2\pi-2\arg\bigl(1-s+se^{j\vartheta_1}\bigr).
\end{aligned}
\label{eq:horo}
\end{equation}
\end{drule}
A cell with loss, several channels or cross-polarization is not an automorphism; its trajectory is a shortened passive image of the horocycle, and Rule~\ref{rule:phase} still bounds it. The amplitude--phase coupling of varactor cells, commonly fitted with empirical models \cite{Abeywickrama2020}, is therefore geometry: the varactor's own line in the impedance plane forces a horocycle, and a horocycle that wraps the origin must come close to it on the far side (\SN{9}). Section~\ref{sec:cases} shows that \eqref{eq:horo} also tells the designer how much amplitude a better cell can recover.

\subsection{Two-State Devices: Switches, Bits, Steps and Filters}
All two-state functions follow from \eqref{eq:twostate} and \eqref{eq:dist}. If the element has CQF $K$, i.e.\ switching ratio $\kappa$ \eqref{eq:CQF}, between the two states used, every pair of responses must satisfy
\begin{equation}
|t_1-t_2|^2\le\frac{\kappa^2}4\,(1-|t_1|^2)(1-|t_2|^2).
\label{eq:master2}
\end{equation}
\begin{drule}[Two-state functions]\label{rule:two}
With power loss ratios $L_{1,2}=|t_{1,2}|^{-2}$:
\begin{enumerate}
\item SPST switch or attenuator step (phase free): $\kappa^2\ge4(\sqrt{L_2}-\sqrt{L_1})^2/[(L_1-1)(L_2-1)]$; with perfect isolation, $|t_{\mathrm{on}}|\le r_{\mathrm c}=(K-1)/(K+1)$, i.e.\ $\IL_{\mathrm{on}}\ge20\log_{10}[(K+1)/(K-1)]$\,dB.
\item Phase bit of step $\Delta\phi$ at equal loss $|t|=e^{-\ell}$: $\sinh\ell\ge2\sin(\Delta\phi/2)/\kappa$; with $\kappa\approx f_c/f$, $\IL\ge8.686\asinh(2\sin\frac{\Delta\phi}{2}\frac f{f_c})$\,dB. For $180^\circ$, $|t|\le(\sqrt K-1)/(\sqrt K+1)$.
\item Tunable or switched filter at frequency $f_0$, from passband loss $L_{\mathrm p}$ to rejection $L_{\mathrm s}$: item~1 with $(L_1,L_2)=(L_{\mathrm p},L_{\mathrm s})$ and the element's $\kappa$ at $f_0$, for every $f_0$ of the tuning range.
\end{enumerate}
At a given frequency each bound is attained, for the specified pair of states, by an ideal $90^\circ$ hybrid with two identical arms, each a lossless reciprocal two-port terminating the element; the two arms hold two physical elements with the budget of one (\SN{10}).
\end{drule}
For $180^\circ$, item~2 is exactly Vendik's reflection-bit limit \cite[eq.~(9)]{Vendik2000}; for any step it reduces, to first order, to Vendik's $|t|=1-2\sin(\Delta\phi/2)/\sqrt K$ \cite[eq.~(11)]{Vendik2000} and to Hines' dissipation ratio $4(f/f_c)\sin(\Delta\phi/2)$ \cite{Hines1964,White1974}. The rule adds that these values bound lossy and nonreciprocal networks too, not only the circuits analyzed there. At $f_c/f=50$ the floors are $0.35$, $0.25$ and $0.13$\,dB for $180^\circ$, $90^\circ$ and $45^\circ$ bits, while a perfectly isolating SPST needs only $0.007$\,dB. A $1$-bit metasurface cell is a $180^\circ$ reflection bit and obeys $\IL\gtrsim17.4f/f_c$\,dB, the value White tabulates as about $18f/f_c$\,dB for one-bit diode shifters \cite{White1974}. A filter that passes at $1$\,dB and rejects $30$\,dB at the same frequency needs $\kappa\ge3.79$ ($K\ge16.3$), i.e.\ $\Delta X/R_s\ge3.79$ for a varactor at that frequency (\SN{13}).

\subsection{Routing Switches and Polarization Switching}
Sending a signal to one of two orthogonal outputs is harder than turning it on and off.
\begin{drule}[Routing]\label{rule:route}
If one fixed input must emerge, in different states, in output states with overlap $c=|\langle\psi_1,\psi_2\rangle|$ and efficiency at least $\eta$, the tuner settings must be at least
\begin{equation}
\delta_{\eta,c}=2\acosh\frac{1-\eta c}{1-\eta}
\label{eq:route}
\end{equation}
apart. For an SPDT, SP$n$T or T/R switch with ideal isolation ($c=0$) built from switches of CQF $K$, toggled singly or in complementary sets,
\begin{equation}
1-\eta\ \ge\ \sqrt{1-r_{\mathrm c}^2}=\frac{2\sqrt K}{K+1}\ \approx\ \frac{2f}{f_c}.
\label{eq:spdt}
\end{equation}
\end{drule}
The SPDT floor scales as $f/f_c$ and the SPST floor as $(f/f_c)^2$; at $f_c/f=50$ they are $0.18$ and $0.007$\,dB. The textbook quarter-wave shunt SPDT, with its impedance level set by a lossless transformation, has loss $R_{\mathrm{on}}/Z+\omega^2C_{\mathrm{off}}^2R_{\mathrm{off}}Z\ge2f/f_c$ and meets \eqref{eq:spdt} to leading order (\SN{11}). The same rule governs polarization switching: right- and left-hand circular patterns of the same shape are orthogonal, so switching between them at $90\%$ total efficiency requires $d\ge5.99$, i.e.\ $K\ge398$ ($\kappa^2\ge396$); switching between linear and circular polarization ($c=1/\sqrt2$) at the same efficiency requires $d\ge3.93$.

\subsection{Attenuators and Complex-Weight Control}
\begin{drule}[Amplitude and complex weights]\label{rule:amp}
An attenuator from loss $\ell_1$ to $\ell_2$ needs
\begin{equation}
\Lambda\ge\ln\frac{\tanh(\ell_2/2)}{\tanh(\ell_1/2)}.
\label{eq:atten}
\end{equation}
A vector modulator that reaches every complex weight with $|t|\le\rho$ needs two continuous controls, with $\Lambda_1\Lambda_2\ge16\rho^2/(1-\rho^2)$.
\end{drule}
The radial path is a geodesic, so \eqref{eq:atten} is also the endpoint distance and is attained. The low-loss end is expensive: a $0.5$--$30$\,dB attenuator needs $\Lambda\ge3.49$, and the requirement diverges as $\ell_1\to0$. The vector-modulator bound counts how many tolerance disks cover the target disk (\SN{12}).

\subsection{Impedance Tuners and Tunable Matching Networks}
\begin{drule}[Reach and coverage]\label{rule:match}
(a) Starting from presented impedance $Z_A=R_A+jX_A$, a tuner of budget $\Lambda$ presents only impedances in
\begin{equation}
(R-R_A\cosh\Lambda)^2+(X-X_A)^2\le R_A^2\sinh^2\Lambda .
\label{eq:reach}
\end{equation}
(b) A presented reflection at distance $\varepsilon$ from the conjugate of the load costs $20\log_{10}\cosh(\varepsilon/2)$\,dB.
(c) Matching every load out to VSWR $S$ within mismatch loss $\mathrm{ML}$\,dB needs at least
\begin{equation}
M\ge\frac{(S-1)^2}{4S\,(10^{\mathrm{ML}/10}-1)}
\label{eq:Mstates}
\end{equation}
switched states. To present every such impedance exactly, two continuous elements need $\Lambda_1\Lambda_2\ge4(S-1)^2/S$, and one continuous element cannot do it for any target disk of nonzero radius; with a finite mismatch allowance the continuous requirement is weaker (\SN{12}).
\end{drule}
An antenna tuner covering VSWR $10$ within $0.1$\,dB needs at least $87$ states; within VSWR $3$ and $0.5$\,dB, at least $3$. The count holds for switched-capacitor banks, MEMS arrays and stub tuners alike \cite{deMingo2004,vanBezooijen2010}.

\subsection{Reconfigurable Antennas and Metasurfaces}
Expanding the radiated field of a passive antenna in power-normalized spherical modes makes the vector of mode amplitudes $y$ a contractive response, with $\|y\|^2$ the total efficiency and $\psi=y/\|y\|$ the normalized pattern; the tuning elements are the tuner and the radiating structure is the network (\SN{14}).
\begin{drule}[Pattern, beam and polarization]\label{rule:antenna}
(a) Switching between patterns with far-field correlation magnitude $c$ at total efficiency $\eta$ requires settings $\delta_{\eta,c}$ of \eqref{eq:route} apart.
(b) Steering continuously at total efficiency at least $\eta$ requires
\begin{equation}
\Lambda\ge2\sqrt{\frac\eta{1-\eta}}\,L_{\mathrm{FS}}\ge2\sqrt{\frac\eta{1-\eta}}\arccos c,
\label{eq:steer}
\end{equation}
with $L_{\mathrm{FS}}$ the Fubini--Study length of the pattern trajectory \cite{Bengtsson2006}.
\end{drule}
For end patterns with $c=0.3$ at $\eta=0.8$, (a) requires $d\ge4.02$ and (b) $\Lambda\ge5.06$. The shortest route between two efficient patterns dips through lower efficiency, where changing the pattern is cheaper; holding the efficiency high throughout costs a length that grows as $(1-\eta)^{-1/2}$, whereas efficient endpoints alone cost about $2\ln[2/(1-\eta)]$ for orthogonal patterns (\SN{14}). Metasurface, reflectarray and transmitarray cells follow Rules~\ref{rule:phase} and~\ref{rule:two} with the Floquet response as the selected channel, and single-channel lossless reflecting cells also follow Rule~\ref{rule:horo}; mutual coupling and fixed losses belong to the network \cite{Gradoni2021}.

\subsection{Multiport Networks, Arrays, State Counts and Cascades}
\begin{drule}[Multiport]\label{rule:multi}
(a) A multiport whose selected block is $e^{-\ell}U$, $U$ unitary, needs $\Lambda\ge L_U/\sinh\ell_{\max}$, $L_U=\int\|U^\dagger\dd U\|_2$; two equal-loss settings are $d=2\asinh[\sin(\Theta/2)/\sinh\ell]$ apart, with $\Theta=\max_i|\operatorname{Arg}\lambda_i(U_0^\dagger U_1)|\in[0,\pi]$ the largest principal eigenphase magnitude.
(b) The number of tuner ports must be at least $\rank(S_1-S_0)$ for any two required states.
(c) In an array with independent phase shifters, each element obeys Rule~\ref{rule:phase} for its own phase travel.
\end{drule}
A reciprocal $2\times2$ switch matrix toggling between bar and in-phase cross states has $\rank\Delta S=2$; bar to quadrature cross has $\rank\Delta S=4$ and cannot be built with fewer than four tuner ports, however they are driven (\SN{15}).

\begin{drule}[States and cascades]\label{rule:states}
(a) One control path of length $\Lambda$ supports at most $1+\lfloor\Lambda/\delta\rfloor$ responses pairwise $\delta$ apart; $m$ independent elements with geodesic ranges $\Lambda_k$ support at most $\prod_k(1+\lfloor\Lambda_k/\delta\rfloor)$.
(b) If fixed stages map their whole input range into radii $r_j<1$, the delivered length is at most $\prod_jr_j$ times the input length.
\end{drule}
For reflection responses, \eqref{eq:mismatch} turns a separation $\delta$ into a mismatch efficiency $\sech^2(\delta/2)$ between a source designed for one state and a load in the other. For patterns or codewords it is only a geometric measure of separation, and a codebook should be scored with its own error metric (\SN{16}).

\begin{table}[!t]
\centering
\caption{Design Requirements for Passive Networks. Here $K$ Is the Element CQF, $\kappa=\sqrt K-1/\sqrt K$, and $\Lambda$ Is Its Path Budget at the Operating Frequency}
\label{tab:summary}
\renewcommand{\arraystretch}{1.25}
\footnotesize
\begin{tabular}{@{}>{\raggedright\arraybackslash}p{0.30\columnwidth}>{\raggedright\arraybackslash}p{0.64\columnwidth}@{}}
\hline
Function & Requirement\\
\hline
Analog phase shifter & $\Lambda\ge\Phi/\sinh\ell_{\max}$\\
 & $\mathrm{FoM}\le6.6\Lambda\sinh\ell_{\max}/\ell_{\max}$\\
 & $N>\Phi/2\pi$ elements (common monotone control, linear-fractional elements)\\
True-time delay, varactor & $\Delta\tau f_H^2\le f_{cv}(1-1/n)\sinh\ell_{\max}/2\pi$\\
Single-varactor cell & horocycle law \eqref{eq:horo}\\
Phase bit $\Delta\phi$ & equal loss: $\sinh\ell\ge2\sin(\Delta\phi/2)/\kappa$; unequal loss: \eqref{eq:master2}\\
SPST, step, filter & $\kappa^2\ge4(\sqrt{L_2}-\sqrt{L_1})^2/[(L_1-1)(L_2-1)]$\\
SPDT, SP$n$T, T/R & ideal isolation, identical switches: $1-\eta\ge2\sqrt K/(K+1)$\\
Attenuator & $\Lambda\ge\ln[\tanh(\ell_2/2)/\tanh(\ell_1/2)]$\\
Vector modulator & exact coverage: $\Lambda_1\Lambda_2\ge16\rho^2/(1-\rho^2)$\\
Impedance tuner & reach \eqref{eq:reach}; switched states \eqref{eq:Mstates}; exact coverage $\Lambda_1\Lambda_2\ge4(S-1)^2/S$\\
Pattern switching & $d\ge2\acosh[(1-\eta c)/(1-\eta)]$\\
Beam steering & $\Lambda\ge2\sqrt{\eta/(1-\eta)}L_{\mathrm{FS}}$\\
Multiport & $\Lambda\ge L_U/\sinh\ell_{\max}$; tuner ports $\ge\rank\Delta S$\\
\hline
\end{tabular}
\end{table}

\section{The $\Lambda$-Chart}\label{sec:chart}
\begin{figure*}[!t]
\centering
\includegraphics[width=\textwidth]{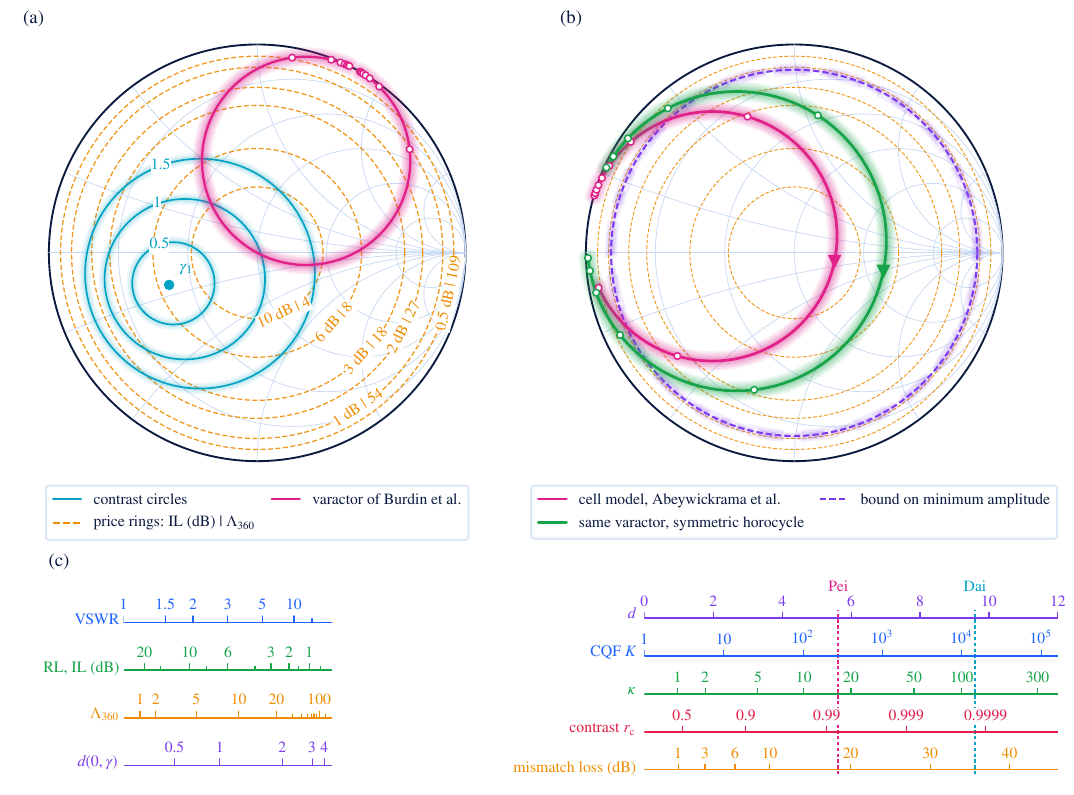}
\caption{The $\Lambda$-chart. (a) Anatomy: Smith grid; price rings (dashed) labeled with insertion loss and full-turn price $\Lambda_{360}$; contrast circles $d=0.5,1,1.5$ about a state $\gamma_1$; and the varactor trajectory of \cite{Burdin2015}, re-centered by a lossless network, with a budget tick every $\Delta\Lambda=10$ ($\Lambda=147.6$). (b) Worked use: reflection of the single-varactor cell model of \cite{Abeywickrama2020}, a horocycle with ticks every $\Delta\Lambda=5$ and worst amplitude $0.20$ (triangle); the symmetric horocycle of equal length that the same varactor reaches through another lossless cell, worst amplitude $0.44$; and the upper bound $0.88$ on the minimum amplitude that any passive cell can keep over this phase travel; individual states may lie outside that circle. (c) Rulers. Left, scaled by the chart radius: VSWR, return or insertion loss, full-turn price and distance from the center; a radius is transferred from the chart with dividers. Right, scaled by the distance $d$ between two states, which is carried over from the left distance ruler: CQF $K$, switching ratio $\kappa$, contrast $r_{\mathrm c}$ and mutual mismatch loss; dashed lines mark the element distances of \cite{Pei2021} and \cite{Dai2020}.}
\label{fig:chart}
\end{figure*}

The Smith chart turned impedance algebra into geometry. Because the budget lives in the same disk, the Smith chart can be extended into a chart for reconfiguration. The $\Lambda$-chart (Fig.~\ref{fig:chart}) keeps the reflection disk and a faint Smith grid, so impedances are plotted exactly as engineers already plot them, and adds four layers.

\emph{Price rings.} Circles $|t|=\rho$, labeled with insertion loss and with $\Lambda_{360}=4\pi\rho/(1-\rho^2)$, the budget of a full constant-loss turn. A phase shifter specification is a ring; the price of $\Phi$ on that ring is $\Lambda_{360}\Phi/2\pi$.

\emph{Budget ticks.} Element trajectories are drawn with a tick every fixed increment of $\Lambda$, computed from \eqref{eq:metric1}. Counting ticks reads the budget, as counting wavelengths reads line length on a Smith chart.

\emph{Contrast circles.} The states within distance $d$ of a state $\gamma_1$ fill a Euclidean disk with center $(1-r^2)\gamma_1/(1-r^2|\gamma_1|^2)$ and radius $r(1-|\gamma_1|^2)/(1-r^2|\gamma_1|^2)$, $r=\tanh(d/2)$. After re-centering on $\gamma_1$, their boundary circles show every position a passive circuit can give the other state relative to the first.

\emph{Rulers.} Below the disk, as on the Smith chart, two sets of scales convert readings into the quantities of Section~\ref{sec:rules}. The left set is scaled by radius: VSWR, return or insertion loss, $\Lambda_{360}$ and the distance $d(0,\gamma)$ from the center. The right set is scaled by the distance $d$ between two states: $K$, $\kappa$, $r_{\mathrm c}$ and the mutual mismatch loss $20\log_{10}\cosh(d/2)$. No ruler belongs to a particular device; a switch, bit or router floor follows from $d$ through Table~\ref{tab:summary}. A contrast radius reaches the right set through the distance ruler; for high-contrast pairs, too close to the rim to read, $d=2\asinh(\kappa/2)$ is computed from the impedances and entered directly.

The chart is used with the operation that makes the Smith chart useful, applied to the budget instead of the impedance: \emph{re-centering}. Any lossless network can move any state to the center; on paper this is the M\"obius map $\gamma\mapsto(\gamma-\gamma_1)/(1-\gamma_1^*\gamma)$, which takes $\gamma_1$ to the center and leaves the contrast of every other state as its radius. Three procedures cover most work (Procedures A, B and D of \SN{17}): (i) for a continuous function, draw the element trajectory with ticks, draw the specification ring, and compare tick count with ring price; (ii) for two states, compute or construct the re-centered radius and read it on the rulers; (iii) for a tuner, draw the coverage disk of radius $\ln S$ and count tolerance disks. Fig.~\ref{fig:chart} shows the first two, and the cell design of Rule~\ref{rule:horo}, on the devices of Section~\ref{sec:cases}; a printable chart and a step-by-step solution of the cell example on it are Figs.~S1--S5 of the Supplementary Information.

\begin{figure*}[!t]
\centering
\includegraphics[width=\textwidth]{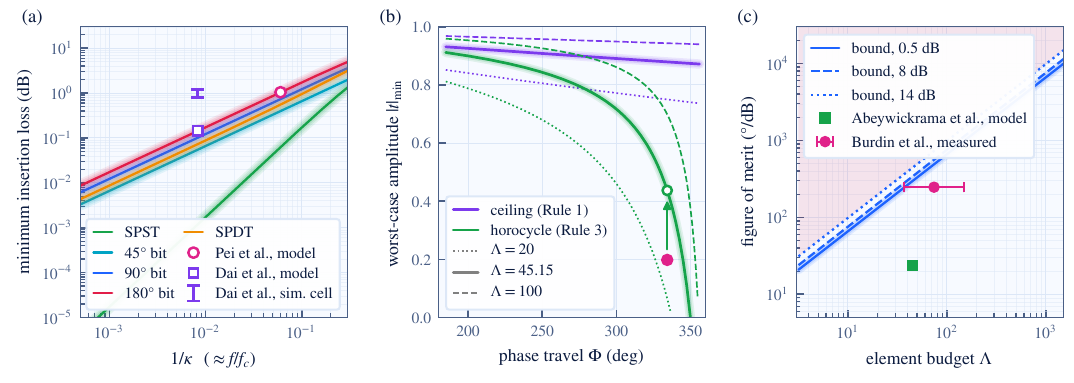}
\caption{Limits and published-device comparisons. (a) Insertion-loss floors of two-state devices versus $1/\kappa$, which approximates $f/f_c$ when reactance dominates: SPST with ideal isolation, $45^\circ$, $90^\circ$ and $180^\circ$ bits, and SPDT. Circle: balanced $180^\circ$ floor of the modeled varactor states of \cite{Pei2021}, $1.04$\,dB. Square: balanced $180^\circ$ floor of the modeled PIN diode of \cite{Dai2020}, $0.14$\,dB; the bar is the simulated cell loss of $0.8$--$1.2$\,dB reported there. (b) Worst-case amplitude versus phase travel for one varactor of budget $\Lambda=20$, $45.15$ and $100$: upper bound for any passive cell (Rule~\ref{rule:phase}) and symmetric horocycle of a lossless cell (Rule~\ref{rule:horo}). Filled circle: the circuit model of \cite{Abeywickrama2020} ($\Lambda=45.15$, $334^\circ$, $0.20$); open circle and arrow: the amplitude $0.44$ that the same varactor reaches. (c) Phase-shifter figure of merit versus element budget: bound $6.6\Lambda\sinh\ell_{\max}/\ell_{\max}$ for peak losses of $0.5$, $8$ and $14$\,dB; the shaded region is excluded for any device whose peak loss is at most $14$\,dB. Bar: measured device of \cite{Burdin2015} over its budget interval $37.2\le\Lambda\le147.6$, circle at the geometric mean $74.1$; its left end is the required budget, on the bound by construction. Square: the circuit model of \cite{Abeywickrama2020} at its $14$\,dB peak loss.}
\label{fig:limits}
\end{figure*}

\section{Relation to Known Limits}\label{sec:prior}
The rule does not replace the classical limits; it explains where they sit. Table~\ref{tab:prior} lists the correspondences, derived in \SN{3}, \SN{7}, \SN{10} and \SN{20}.

\emph{Two-state limits.} Hines showed, by perturbation, that a diode phase bit dissipates at least the fraction $4(f/f_c)\sin(\Delta\phi/2)$, with the switching cutoff frequency $f_c=1/(2\pi C_{\mathrm{off}}\sqrt{R_{\mathrm{on}}R_{\mathrm{off}}})$ \cite{Hines1964,White1974}; Kawakami proved that a two-state one-port has a figure of merit unchanged by lossless reciprocal networks \cite{Kawakami1965}; Vendik \emph{et al.} defined the CQF and derived the loss of switches and digital phase shifters from it \cite{Vendik2000}. In the present framework these are the statement that the contrast \eqref{eq:contrast} is invariant under lossless networks: Kawakami's invariant is $r_{\mathrm c}$, Vendik's CQF is $K=e^d$, and Hines' $f_c/f$ approximates $\kappa=2\sinh(d/2)$ when reactance dominates. Rule~\ref{rule:two} reproduces Vendik's $180^\circ$ limit exactly and Hines' loss to first order. Theorem~\ref{thm:main} adds that lossy and nonreciprocal networks can only reduce the contrast, extends the result to multiports as a nonexpansion, and supplies the attaining circuit of Rule~\ref{rule:two}. White's analysis of diode phase shifters for arrays \cite{White1974} and the RF-MEMS switch and phase-shifter literature \cite{Rebeiz2003,Hayden2003} report figures of merit that the same inequalities bound.

\emph{Phase-shifter figures of merit.} The ratio $^\circ$/dB, the material figure of merit $\tau/\tan\delta$ of liquid crystals \cite{Jakoby2020} and the CQF of ferroelectric capacitors \cite{Vendik2000,Tagantsev2003} are established figures of merit, and the CQF work already derives loss limits for switches and phase bits. Equations \eqref{eq:fom} and \eqref{eq:material} turn them into a bound: the device figure of merit is at most $6.6\Lambda\sinh\ell_{\max}/\ell_{\max}$, i.e.\ $6.6$ times the element budget at low loss, and for a tunable dielectric with constant loss tangent and small $\tau$ the budget reduces to the material figure of merit. Jakoby \emph{et al.} give the optimum figure of merit of a liquid-crystal phase shifter as $\mathrm{FoM}_{\mathrm{opt}}\approx6.6\,\tau/\tan\delta_{\max}$ $^\circ$/dB \cite{Jakoby2020}; for constant $\tan\delta$, $\Lambda\approx\tau/\tan\delta$ and \eqref{eq:fom} gives the same number, otherwise the bound is $6.6\Lambda$ from \eqref{eq:material}, and Theorem~\ref{thm:main} shows that it bounds every topology.

\emph{Broadband matching and antenna limits.} Bode, Fano and Youla limit how well a fixed network can match a fixed load over a band \cite{Bode1945,Fano1950a,Fano1950b,Youla1964}; Chu, Wheeler and Gustafsson \emph{et al.} limit the bandwidth of small antennas \cite{Chu1948,Wheeler1947,Gustafsson2007}. These constrain frequency behavior at a fixed state; the present rule constrains how far the state can move at each frequency. The two kinds of limit are independent and both apply.

\emph{RIS and metasurface models.} Empirical amplitude--phase models of RIS cells \cite{Abeywickrama2020} and impedance-based RIS models with mutual coupling \cite{Gradoni2021} describe the response of given cells. The horocycle law (Rule~\ref{rule:horo}) derives the coupling from the varactor line, and Rule~\ref{rule:phase} gives the ceiling that no cell design can pass. Electromagnetically consistent bounds on realizing targeted MIMO operators with RIS and dynamic metasurface antennas \cite{DelHougne2026b} and prototype-aware bounds for programmable metasurfaces \cite{DelHougne2026} optimize over a given structure, including mutual coupling and quantized states; the present budget is computed from the tuning element alone, before the structure exists, and bounds every structure that could be built around it.

\emph{Optical and pole--zero limits.} Bounds on optical response from material loss \cite{Miller2016,Shim2024}, on graphene modulators \cite{Tamagnone2014} and on phase and amplitude actuators \cite{Zanotto2015} use material and electromagnetic information; constant-amplitude $2\pi$ phase can be engineered through pole--zero winding \cite{Krasnok2026}. That work synthesizes the response with several controls; here the tunable region enters only through its scattering response. Rule~\ref{rule:phase} bounds the budget of the winding along any control path; Rule~\ref{rule:count} adds an element count for elements linear-fractional in one common monotone control.

\emph{Geometry.} The hyperbolic geometry of the Smith chart \cite{Ohira2020,Muller2011} and its operator generalization \cite{Helton1982,Harris1979} are classical; the $\Lambda$-chart adds the budget, price and contrast layers that turn that geometry into a reconfiguration tool.

\begin{table}[!t]
\centering
\caption{Classical Limits as Special Cases}
\label{tab:prior}
\renewcommand{\arraystretch}{1.2}
\footnotesize
\begin{tabular}{@{}>{\raggedright\arraybackslash}p{0.36\columnwidth}>{\raggedright\arraybackslash}p{0.56\columnwidth}@{}}
\hline
Classical result & Form in this work\\
\hline
CQF invariance \cite{Kawakami1965,Vendik2000} & contrast $r_{\mathrm c}$ invariant under lossless networks; $K=e^d$, $r_{\mathrm c}=\tanh(d/2)$\\
Diode phase-shifter loss \cite{Hines1964,White1974,Vendik2000} & $\sinh\ell\ge2\sin(\Delta\phi/2)/\kappa$, attained\\
FoM $^\circ$/dB & $\le6.6\Lambda\sinh\ell_{\max}/\ell_{\max}$\\
LC material FoM $\tau/\tan\delta$ \cite{Jakoby2020} & $\Lambda$ of a tunable dielectric for small $\tau$\\
Smith chart \cite{Smith1939} & re-centering $=$ disk automorphism; $\ln\mathrm{VSWR}=d(0,\gamma)$\\
Bode--Fano \cite{Bode1945,Fano1950a,Fano1950b} & complementary: bandwidth at fixed state\\
RIS amplitude--phase models \cite{Abeywickrama2020} & horocycle law\\
\hline
\end{tabular}
\end{table}

\section{Case Studies of Published Devices}\label{sec:cases}
Four published microwave case studies, analyzed in full in \SN{18}, show the rule at work (Table~\ref{tab:cases}, Fig.~\ref{fig:limits}).

\emph{Analog reflection-type phase shifter} \cite{Burdin2015}. Two $\pi$-network arms with three jointly biased varactors each give a measured $385^\circ$ at $2$\,GHz with insertion loss $0.98\pm0.58$\,dB, i.e.\ at most $1.56$\,dB. The varactor model ($R_s=1\,\Omega$, $0.45$\,pF at $20$\,V to $2.72$\,pF at $0$\,V; the measurement used $0.5$--$20$\,V) gives $\Lambda\le147.6$; the six identical diodes count once. The specification requires $\Lambda\ge37.2$, so the design uses at least $25\%$ of its budget, and its reported $246^\circ$/dB lies below the bound $979^\circ$/dB. Rule~\ref{rule:count} demands at least two elements for $385^\circ$. Equation~\eqref{eq:ceiling} sets a ceiling of $7.9$\,GHz for this diode and specification, and at $2$\,GHz it excludes any series resistance above $3.97\,\Omega$ for the same capacitance range.

\emph{Single-varactor cell model} \cite{Abeywickrama2020,Zhu2013}. Abeywickrama \emph{et al.} \cite{Abeywickrama2020} model a single-varactor cell ($R=2.5\,\Omega$, $C=0.47$--$2.35$\,pF, lossless surroundings) at $2.4$\,GHz, after the metasurface of Zhu \emph{et al.} \cite{Zhu2013}. The varactor gives $\Lambda=45.15$. The computed cell trajectory has exactly this length, as a lossless two-port requires, since it acts as a disk automorphism (\SN{2}), and spans $334^\circ$ with a minimum amplitude of $0.20$, matching the amplitude model fitted in \cite{Abeywickrama2020}. The horocycle law shows that the same varactor, placed symmetrically by a different lossless cell, reaches a minimum amplitude of $0.44$ over the same phase range, and Rule~\ref{rule:phase} shows that no passive cell can hold more than $0.88$ throughout. The improved cell is explicit: from the varactor outward, a $102.7^\circ$ line, an ideal transformer that lowers the impedance $62.05$ times toward the port, and an $89.8^\circ$ line, all at $377\,\Omega$, give a minimum amplitude of $0.437$ over $334.2^\circ$ with the same length $45.15$ (\SN{18}). With the same varactor, an ideal lossless cell thus more than doubles the worst-case amplitude at the design frequency, and the horocycle law gives that cell.

\emph{One-bit varactor RIS} \cite{Pei2021}. The two simulated design states ($7$ and $16$\,V) of the SMV2019 varactor model at $5.8$\,GHz have $K=278$ ($\kappa^2=276$), contrast $r_{\mathrm c}=0.993$; the measured $180^\circ$ pair, near $3.8$ and $16$\,V, lies on the same sweep. Any $180^\circ$ cell built with them, with equal amplitudes in its two states, loses at least $1.04$\,dB in each; over the $0$--$16$\,V sweep the model gives $\Lambda\le30.4$ for any monotone $C(V)$ and $R(V)$ between the tabulated biases, so the measured $250^\circ$ range costs at least $1.24$\,dB phase-weighted mean loss; linear interpolation of the model gives $\Lambda\approx25.9$ and an estimate of $1.46$\,dB.

\begin{figure}[!t]
\centering
\includegraphics[width=\columnwidth]{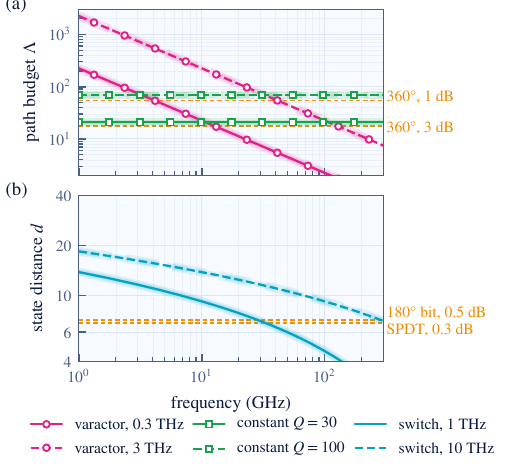}
\caption{Available tuning resources versus frequency. (a) Path budget $\Lambda$ of continuous tuners: varactors with $f_{cv}=0.3$ and $3$\,THz and $n=4$, and constant-$Q$ dielectrics with $Q=30$ and $100$ and $n=2$; the labeled horizontal requirement lines are for $360^\circ$ of phase at $1$ and $3$\,dB peak loss. (b) Distance $d$ between the two states of switches with $f_c=1$ and $10$\,THz and $R_{\mathrm{off}}=R_{\mathrm{on}}$, so that $\kappa=f_c/f$; the labeled horizontal requirement lines are for a $180^\circ$ bit at $0.5$\,dB and an SPDT at $0.3$\,dB. Each crossing is a frequency ceiling for the stated element model in any passive topology.}
\label{fig:tech}
\end{figure}
\emph{Two-bit PIN RIS} \cite{Dai2020}. The SMP1340 model at $2.3$\,GHz has $K\approx\kappa^2=1.47\times10^4$, giving balanced-pair floors of $0.14$\,dB for $180^\circ$ and $0.10$\,dB for $90^\circ$ steps. The simulated state losses of $0.8$--$1.2$\,dB reported for this design are far above them. Because the states are not balanced, the loss of one state above the floor cannot be attributed to extra dissipation in that state; apportioning the loss between the fixed structure and the diode's embedding needs a separate loss analysis.

\begin{table}[!t]
\centering
\caption{Case Studies (Details in \SN{18})}
\label{tab:cases}
\renewcommand{\arraystretch}{1.2}
\footnotesize
\begin{tabular}{@{}>{\raggedright\arraybackslash}p{0.19\columnwidth}>{\raggedright\arraybackslash}p{0.17\columnwidth}>{\raggedright\arraybackslash}p{0.23\columnwidth}>{\raggedright\arraybackslash}p{0.25\columnwidth}@{}}
\hline
Device & Budget & Requirement or floor & Finding\\
\hline
RTPS \cite{Burdin2015} & $\Lambda\le147.6$ & $\Lambda\ge37.2$ & $\ge25\%$ used; ceiling $7.9$\,GHz\\
Cell model \cite{Abeywickrama2020} & $\Lambda=45.15$ & horocycle, $334^\circ$ & $|t|_{\min}$: $0.20\to0.44$ possible\\
1-bit RIS \cite{Pei2021} & $K=278$ & $\ge1.04$\,dB, balanced pair & diode floor; sweep $\ge1.24$\,dB mean\\
2-bit RIS \cite{Dai2020} & $K=1.47\!\times\!10^4$ & $\ge0.14$\,dB, balanced pair & simulated states far above floor\\
\hline
\end{tabular}
\end{table}

\section{Design Procedure}\label{sec:procedure}
\begin{enumerate}
\item Draw the tuner boundary around everything that changes with the control.
\item Obtain its trajectory $Z(q)$ or $\Gamma(q)$ at the design frequencies from a datasheet, a device model or VNA data, and compute $\Lambda$ or $K$ (Table~\ref{tab:budgets}).
\item Translate the specification into requirements (Table~\ref{tab:summary}): path length for continuous operations, distance for state pairs, rank and element count for structure.
\item If a requirement exceeds an upper bound on the budget at any frequency of the band, the specification is excluded for this element model; change the element, the frequency or the specification.
\item If all requirements pass, the ratio $u=\Lambda_{\mathrm{req}}/\Lambda$ is the budget utilization the topology must reach, a single number for comparing designs that use different elements.
\end{enumerate}
Fig.~\ref{fig:tech} compares the resources of the main tuning technologies, path length $\Lambda$ for continuous tuning and state distance $d$ for switching, with representative requirements; where a resource falls below its requirement, the specification is excluded for that element model in any passive topology.

\section{Scope and Beyond Microwaves}\label{sec:scope}
The rules are necessary conditions for a fixed linear passive network. Passing them means the budget does not rule out the design; circuit synthesis, bandwidth, power handling, linearity, switching speed and bias design remain separate tasks. Bandwidth is complementary: the rules apply at each frequency, the worst frequency decides, and Bode--Fano theory limits matching across the band. Ideal lossless elements have infinite budget and are not constrained; a real element has a finite budget when it moves smoothly over a finite control range and its loss stays above zero in every state, which keeps its trajectory away from the lossless boundary. Large-signal operation, time-modulated circuits whose complete operator is not passive, and active circuits are outside the theorem, which then applies to the passive core between active stages (\SN{20}).

Theorem~\ref{thm:main} uses only power-normalized amplitudes and a structure that does not create energy, not voltages, currents or fields. The same rule therefore governs tunable photonic circuits, optical and acoustic metasurfaces, acoustic loads and passive single-photon processing, where $\|y\|^2$ is the survival probability of the photon (\SN{19}).

\section{Conclusion}
A tuning element arrives with a finite budget, and the circuit around it decides how much of that budget reaches the output, never how large it is. The budget is computed from the element alone; its unit is the hyperbolic distance of the Smith chart; and it reduces to the VSWR, the mismatch loss and the commutation quality factor in their familiar forms. Simple circuits attain the two-state and attenuation bounds, a constant-loss trajectory attains the phase bound, and the $\Lambda$-chart makes the limits as easy to read as impedance on a Smith chart. The practical consequence is a change in the order of design: check the element against the specification first, then optimize the topology for utilization.

\section*{Software Availability}
The $\Lambda$-Lab software (version 1.0.0) is available at \url{https://doi.org/10.5281/zenodo.23942414}.

\end{document}


\maketitle
\begingroup\small\tableofcontents\endgroup
\bigskip

\noindent\textbf{How to read this document.} Notes S1--S5 build the geometry and prove the theorem from first principles; each step is written out so that it can be checked line by line. Notes S6--S16 derive every design rule of the main text. Note S17 is a user's manual for the $\Lambda$-chart, and Note S18 solves experimental examples with it. Note S19 transfers the rule to optical, acoustic and quantum waves, Note S20 states the scope conditions, and Note S21 lists the numerical checks. Equation, table and figure numbers carry the prefix S; numbers without the prefix refer to the main text.

\medskip
\noindent\textbf{Symbols.} The main symbols; others are defined where they are used.
\begin{center}
\small
\renewcommand{\arraystretch}{1.15}
\begin{tabular}{@{}>{$}l<{$}>{\raggedright\arraybackslash}p{0.36\textwidth}>{$}l<{$}>{\raggedright\arraybackslash}p{0.36\textwidth}@{}}
\toprule
S & scattering matrix of the fixed network, blocks $\See,\Set,\Ste,\Stt$ (e: external ports, t: tuner ports); in coverage problems $S$ is a VSWR & \Lambda & path budget: hyperbolic length of a trajectory\\
\Gamma,\ \gamma & reflection of the tuner (matrix, scalar) & \Lambda_{360} & price of a full $360^\circ$ turn on a ring of constant amplitude\\
F,\ t & device response (matrix, scalar) & \Phi & phase travel along a trajectory\\
D_X & defect $(I-X^\dagger X)^{1/2}$; $D_X^2$ is the lost power fraction & \ell & loss in nepers, $|t|=e^{-\ell}$; $\IL=8.686\,\ell$ dB\\
Q_\Gamma,\ P_\Gamma & access and return factors of the network, $\le1$ & \rho & radius of a price ring, $|t|=\rho$; $\rho(\gamma_1,\gamma_2)$ is the pseudo-hyperbolic distance\\
Z=R+jX & impedance; $Z_0$ reference impedance & q & control variable (bias, state)\\
d(\cdot,\cdot) & hyperbolic distance; $d(0,\gamma)=\ln\mathrm{VSWR}$ & n & capacitance ratio $C_{\max}/C_{\min}$\\
r_{\mathrm c} & contrast of two states, $\tanh(d/2)$ & Q,\ Q_{\max} & element quality factor, its maximum over the sweep\\
K & CQF of Vendik \emph{et al.}, $e^d$ & R_s & series resistance of a varactor\\
\kappa & switching ratio, $2\sinh(d/2)=|Z_1-Z_2|/\sqrt{R_1R_2}$ & f_c,\ f_{cv} & switch and varactor cutoff frequencies\\
s,\ \vartheta & Euclidean radius and angle parameter of a horocycle (Note~S9) & \tau,\ \tan\delta & tunability and loss tangent of a dielectric\\
\eta,\ \eta_{\mathrm m} & efficiency of a selected output; mismatch efficiency & N,\ M & numbers of elements and of states\\
c & overlap of two normalized patterns or outputs & \varepsilon,\ \varepsilon_0 & tolerance radius (hyperbolic) of coverage\\
\bottomrule
\end{tabular}
\end{center}

\section{Power Waves and Passivity}\label{sec:S1}
At a port with real reference resistance $R_{\mathrm r}>0$, the power waves \cite{Kurokawa1965}
\begin{equation}
a=\frac{V+R_{\mathrm r}I}{2\sqrt{R_{\mathrm r}}},\qquad b=\frac{V-R_{\mathrm r}I}{2\sqrt{R_{\mathrm r}}}
\end{equation}
satisfy $|a|^2-|b|^2=\Re(VI^*)$, the power delivered into the port. For an $n$-port with $b=Sa$ the delivered power is
\begin{equation}
P=\|a\|^2-\|Sa\|^2=a^\dagger(I-S^\dagger S)a .
\end{equation}
\emph{Passive} means $P\ge0$ for every excitation, i.e.\ $S^\dagger S\preceq I$; equivalently the largest singular value obeys $\|S\|_2\le1$, and $S$ is a contraction \cite{Youla1959}. \emph{Lossless} means $S^\dagger S=I$. \emph{Strictly passive} means $\|S\|_2<1$: every excitation loses some power. Reciprocity ($S=S^{\mathsf T}$) is never needed below.

For a contraction $X$ define the defect matrices
\begin{equation}
D_X=(I-X^\dagger X)^{1/2},\qquad D_{X^\dagger}=(I-XX^\dagger)^{1/2}.
\end{equation}
They are positive semidefinite and invertible exactly when $X$ is strict. $D_X^2$ measures the power lost per unit incident power; $D_X\to0$ means losslessness. All norms are operator (spectral) norms unless stated otherwise.

\section{Hyperbolic Geometry of the Reflection Disk}\label{sec:S2}
\subsection{Automorphisms and lossless two-ports}
For $|a|<1$ and real $\theta$ the Möbius map
\begin{equation}
\varphi(\gamma)=e^{j\theta}\frac{\gamma-a}{1-a^*\gamma}
\label{eq:aut}
\end{equation}
maps the unit disk onto itself and satisfies the identity
\begin{equation}
1-|\varphi(\gamma)|^2=\frac{(1-|a|^2)(1-|\gamma|^2)}{|1-a^*\gamma|^2},
\label{eq:autid}
\end{equation}
which follows by expanding $|1-a^*\gamma|^2-|\gamma-a|^2=(1-|a|^2)(1-|\gamma|^2)$. Every analytic bijection of the disk has the form \eqref{eq:aut} \cite{Ahlfors1973}.

\begin{lemma}\label{lem:lossless}
A lossless two-port with $s_{21}\neq0$, terminated at port~2 by a load $\gamma$, presents at port~1 a reflection $\Gamma_{\mathrm{in}}=\varphi(\gamma)$ with $\varphi$ an automorphism. Conversely every automorphism is the input reflection of a lossless reciprocal two-port.
\end{lemma}
\begin{proof}
With $a_2=\gamma b_2$ and $b_2=s_{21}a_1+s_{22}a_2$ we get $b_2=s_{21}a_1/(1-s_{22}\gamma)$ and $\Gamma_{\mathrm{in}}=s_{11}+s_{12}s_{21}\gamma/(1-s_{22}\gamma)$, a Möbius function of $\gamma$. Losslessness gives $|a_1|^2(1-|\Gamma_{\mathrm{in}}|^2)=|b_2|^2(1-|\gamma|^2)$: power entering port~1 equals power absorbed by the load. Hence
\begin{equation}
1-|\Gamma_{\mathrm{in}}|^2=\frac{|s_{21}|^2(1-|\gamma|^2)}{|1-s_{22}\gamma|^2}.
\label{eq:lossless}
\end{equation}
So $|\gamma|<1$ implies $|\Gamma_{\mathrm{in}}|<1$ and $|\gamma|=1$ implies $|\Gamma_{\mathrm{in}}|=1$: the Möbius map sends the disk into itself and its boundary onto the boundary, hence onto the disk, and it is an automorphism. For the converse, a line of electrical length $\beta l$ gives the rotation $\gamma\mapsto e^{-2j\beta l}\gamma$, and an ideal transformer of ratio $N$, with $Z_{\mathrm{in}}=Z_L/N^2$, gives $\gamma\mapsto(\gamma-a)/(1-a\gamma)$ with real $a=(N^2-1)/(N^2+1)$. Writing $a=|a|e^{j\alpha}$, every automorphism \eqref{eq:aut} equals rotation by $e^{j(\theta+\alpha)}$ after the real map $(\cdot-|a|)/(1-|a|\,\cdot)$ after rotation by $e^{-j\alpha}$, i.e.\ line--transformer--line.
\end{proof}

\subsection{Pseudo-hyperbolic distance, metric and Schwarz--Pick}
Define the pseudo-hyperbolic distance
\begin{equation}
\rho(\gamma_1,\gamma_2)=\left|\frac{\gamma_2-\gamma_1}{1-\gamma_1^*\gamma_2}\right| .
\end{equation}
It is invariant under automorphisms. A rotation clearly leaves it unchanged, and every automorphism is a rotation composed with a map $\varphi_a(\gamma)=(\gamma-a)/(1-a^*\gamma)$, so it suffices to check $\varphi_a$. Expanding numerators over the common denominator $(1-a^*x)(1-a^*y)$ gives
\begin{equation}
\varphi_a(y)-\varphi_a(x)=\frac{(1-|a|^2)(y-x)}{(1-a^*x)(1-a^*y)},\qquad
1-\varphi_a(x)^*\varphi_a(y)=\frac{(1-|a|^2)(1-x^*y)}{(1-ax^*)(1-a^*y)} .
\end{equation}
Dividing the first by the second, the factor $1-|a|^2$ cancels and the remaining denominators have equal modulus, so $\rho(\varphi_a(x),\varphi_a(y))=\rho(x,y)$. In particular the map with $a=\gamma_1$ sends $\gamma_1$ to $0$ and $\gamma_2$ to a point of modulus $\rho(\gamma_1,\gamma_2)$. The infinitesimal version is the Poincaré line element and the distance it generates:
\begin{equation}
\dd\Lambda=\frac{2|\dd\gamma|}{1-|\gamma|^2},\qquad d(\gamma_1,\gamma_2)=2\atanh\rho(\gamma_1,\gamma_2).
\label{eq:poincare}
\end{equation}
The factor two gives curvature $-1$ \cite{Beardon1983}. Along the radius, $d(0,\gamma)=2\atanh|\gamma|$, and the diameter is a shortest path.

\begin{theorem}[Schwarz--Pick \cite{Ahlfors1973}]\label{thm:SP}
If $f$ is analytic and maps the open disk into itself, then $\rho(f(\gamma_1),f(\gamma_2))\le\rho(\gamma_1,\gamma_2)$ and $2|f'(\gamma)|/(1-|f(\gamma)|^2)\le2/(1-|\gamma|^2)$, and equality at one pair of distinct points, or at one point, holds only for automorphisms.
\end{theorem}

A passive two-port terminated by $\gamma$ is such a map (Note~S4 proves the matrix version), so the contrast of two states cannot increase. This is the content of Section~II-B of the main text.

\subsection{Circles, horocycles and geodesics}
\emph{Hyperbolic circles.} The set $d(\gamma,\gamma_1)\le D$, with $r=\tanh(D/2)$, is $|\gamma-\gamma_1|^2\le r^2|1-\gamma_1^*\gamma|^2$. Completing the square gives the Euclidean disk
\begin{equation}
|\gamma-c_D|\le R_D,\qquad c_D=\frac{(1-r^2)\gamma_1}{1-r^2|\gamma_1|^2},\qquad R_D=\frac{r(1-|\gamma_1|^2)}{1-r^2|\gamma_1|^2}.
\label{eq:hcircle}
\end{equation}
Its hyperbolic area is $\int_0^{r}8\pi s\,\dd s/(1-s^2)^2=4\pi r^2/(1-r^2)=4\pi\sinh^2(D/2)$, independent of the center.

\emph{Geodesics} are diameters and circular arcs orthogonal to the unit circle. \emph{Horocycles} are circles inside the disk tangent to the unit circle; they are the limits of hyperbolic circles whose center moves to the boundary.

\emph{The Smith grid.} With $\zeta=Z/Z_0=r+jx$ and $\gamma=(\zeta-1)/(\zeta+1)$, the constant-resistance circles have center $r/(1+r)$ and radius $1/(1+r)$; they touch the unit circle at $\gamma=1$ and are horocycles. The constant-reactance circles have center $1+j/x$ and radius $1/|x|$; they pass through $\gamma=1$ at right angles to the unit circle and are geodesics. The Smith chart is therefore drawn with horocycles and geodesics of the same geometry that carries the budget; this is why constant-resistance varactors trace horocycles (Note~S9) and why a pure variable resistance moves along a geodesic.

\subsection{Impedance half-plane}
The Cayley map $Z=Z_0(1+\gamma)/(1-\gamma)$ carries the disk to the right half-plane $\Re Z>0$ and \eqref{eq:poincare} to
\begin{equation}
\dd\Lambda=\frac{|\dd Z|}{\Re Z}
\label{eq:halfplane}
\end{equation}
(Note~S5). The reference $Z_0$ cancels, so the budget of an element does not depend on the characteristic impedance chosen to describe it.

\section{Contrast, Mismatch and the CQF}\label{sec:S3}
\subsection{The basic identity}
For any $x,y$ in the disk,
\begin{equation}
|1-x^*y|^2-(1-|x|^2)(1-|y|^2)=|x-y|^2 .
\label{eq:basic}
\end{equation}
Both sides expand to $|x|^2+|y|^2-2\Re(x^*y)$. Dividing by $|1-x^*y|^2$ gives $1-\rho^2=(1-|x|^2)(1-|y|^2)/|1-x^*y|^2$, and with $\rho=\tanh(d/2)$,
\begin{equation}
\sech^2\frac d2=\frac{(1-|x|^2)(1-|y|^2)}{|1-x^*y|^2},\qquad
\sinh^2\frac d2=\frac{|x-y|^2}{(1-|x|^2)(1-|y|^2)} .
\label{eq:sech}
\end{equation}
The second form is equation~(13) of the main text.

\subsection{VSWR}
$d(0,\gamma)=2\atanh|\gamma|=\ln\frac{1+|\gamma|}{1-|\gamma|}=\ln\mathrm{VSWR}$.

\subsection{Mismatch}
A generator of reflection $\Gamma_{\mathrm s}$ connected to a load $\Gamma_{\mathrm L}$ delivers the fraction \cite{Pozar2012}
\begin{equation}
\eta_{\mathrm m}=\frac{(1-|\Gamma_{\mathrm s}|^2)(1-|\Gamma_{\mathrm L}|^2)}{|1-\Gamma_{\mathrm s}\Gamma_{\mathrm L}|^2}
\end{equation}
of its available power. Setting $x=\Gamma_{\mathrm s}^*$, $y=\Gamma_{\mathrm L}$ in \eqref{eq:sech} gives $\eta_{\mathrm m}=\sech^2[d(\Gamma_{\mathrm s}^*,\Gamma_{\mathrm L})/2]$. The conjugate match $\Gamma_{\mathrm L}=\Gamma_{\mathrm s}^*$ is $d=0$, and the mismatch loss is $-10\log_{10}\eta_{\mathrm m}=20\log_{10}\cosh(d/2)$\,dB. Two states one unit apart therefore mismatch each other by $20\log_{10}\cosh(1/2)=1.04$\,dB.

\subsection{Contrast and CQF}
The contrast of two states is $r_{\mathrm c}=\rho(\gamma_1,\gamma_2)=\tanh(d/2)$. By Lemma~\ref{lem:lossless}, matching state~1 with a lossless network is the automorphism with $a=\gamma_1$; state~2 then sits at radius $r_{\mathrm c}$. In impedances, substituting $\gamma_i=(Z_i-Z_0)/(Z_i+Z_0)$,
\begin{equation}
1-|\gamma_i|^2=\frac{4Z_0R_i}{|Z_i+Z_0|^2},\qquad
1-\gamma_1^*\gamma_2=\frac{2Z_0(Z_1^*+Z_2)}{(Z_1^*+Z_0)(Z_2+Z_0)},
\end{equation}
so \eqref{eq:sech} gives $\sech^2(d/2)=4R_1R_2/|Z_1^*+Z_2|^2$ and
\begin{equation}
\sinh^2\frac d2=\frac{|Z_1^*+Z_2|^2-4R_1R_2}{4R_1R_2}=\frac{(R_1-R_2)^2+(X_1-X_2)^2}{4R_1R_2}=\frac{\kappa^2}4 ,
\end{equation}
which defines the switching ratio $\kappa=|Z_1-Z_2|/\sqrt{R_1R_2}$. Since $|Z_1^*+Z_2|^2-|Z_1-Z_2|^2=4R_1R_2$, the contrast is
\begin{equation}
r_{\mathrm c}^2=1-\sech^2\frac d2=\frac{|Z_1-Z_2|^2}{|Z_1^*+Z_2|^2},\qquad r_{\mathrm c}=\left|\frac{Z_1-Z_2}{Z_1+Z_2^*}\right|,
\end{equation}
which is the invariant of Kawakami \cite{Kawakami1965}. Vendik \emph{et al.} \cite{Vendik2000} transform the pair losslessly into two resistances $R^{(1)}_{\mathrm{in}}<R^{(2)}_{\mathrm{in}}$ and define the CQF as $K=R^{(2)}_{\mathrm{in}}/R^{(1)}_{\mathrm{in}}$. On the positive real axis $|\dd Z|/\Re Z=\dd R/R$, and the real axis is a geodesic, so the distance between the two resistances is $\ln K$. The distance is invariant (Lemma~\ref{lem:lossless}), hence
\begin{equation}
K=e^{d}=\frac{1+r_{\mathrm c}}{1-r_{\mathrm c}},\qquad K+\frac1K-2=4\sinh^2\frac d2=\kappa^2,\qquad \kappa=\sqrt K-\frac1{\sqrt K}.
\end{equation}
The middle identity is eq.~(33) of \cite{Vendik2000}. Also $\cosh(d/2)=\sqrt{1+\kappa^2/4}=(K+1)/(2\sqrt K)$. For a switch with $X_{\mathrm{off}}\gg R_{\mathrm{on}},R_{\mathrm{off}}$, $\kappa\approx f_c/f$ with Hines' cutoff frequency $f_c=1/(2\pi C_{\mathrm{off}}\sqrt{R_{\mathrm{on}}R_{\mathrm{off}}})$ \cite{Hines1964,White1974}, and $K\approx\kappa^2$ when $\kappa\gg1$. The closed form often quoted as the CQF of ferroelectric capacitors, $(n-1)^2/(n\tan\delta_1\tan\delta_2)$, is $\kappa^2$ with the difference of the two loss resistances neglected; it agrees with $K$ only at high contrast, and at low contrast the two must not be mixed (for example $\kappa^2=14.4$ against $K=16.3$ in Note~S13).

\subsection{Ruler formulas}
Table~\ref{tab:rulers} collects the conversions printed on the rulers of the $\Lambda$-chart. Each follows from the notes indicated.
\begin{table}[!ht]
\centering
\caption{Ruler formulas of the $\Lambda$-chart. The left rulers are read from a radius $r$, the right rulers from a distance $d$.}
\label{tab:rulers}
\renewcommand{\arraystretch}{1.3}
\begin{tabular}{lll}
\toprule
Reading of $r$ & Quantity & Formula\\
\midrule
reflection $|\gamma|$ & VSWR, return loss & $(1+r)/(1-r)$, \ $-20\log_{10}r$\\
transmission $|t|$ & insertion loss, full-turn price & $-20\log_{10}r$, \ $\Lambda_{360}=4\pi r/(1-r^2)$ (Note~S7)\\
contrast $r_{\mathrm c}$ & distance, CQF & $d=2\atanh r$, \ $K=(1+r)/(1-r)$\\
distance $d$ & switching ratio & $\kappa=2\sinh(d/2)=\sqrt K-1/\sqrt K$\\
distance $d$ & mutual mismatch loss & $20\log_{10}\cosh(d/2)$ (Note~S3)\\
\bottomrule
\end{tabular}
\end{table}

\section{The Passive Reconfiguration Theorem}\label{sec:S4}
\emph{Plan of the proof.} Three facts are needed. First, a formula for how the response changes when the tuner changes (S4.2). Second, two energy balances, which say that the network can only add loss to the tuner's own loss (S4.3, S4.4). Third, a rearrangement of the first formula in which each factor is measured against the loss that limits it; the energy balances then show that the outer factors are at most one (S4.5). The theorem follows by taking norms.

\subsection{Setting}
The network $S=\bigl[\begin{smallmatrix}\See&\Set\\ \Ste&\Stt\end{smallmatrix}\bigr]$ is a contraction. The tuner $\Gamma$ is strict, $\|\Gamma\|_2<1$, and its input and output dimensions may differ. Then $\|\Stt\Gamma\|_2<1$, so
\begin{equation}
R_\Gamma=(I-\Stt\Gamma)^{-1},\qquad L_\Gamma=(I-\Gamma\Stt)^{-1}
\end{equation}
exist by the Neumann series, and the response is $F(\Gamma)=\See+\Set\Gamma R_\Gamma\Ste$ \cite{Redheffer1960}.

\subsection{Derivative and finite difference}
Two identities are used repeatedly. The first, $\Gamma R_\Gamma=L_\Gamma\Gamma$, follows by multiplying $\Gamma(I-\Stt\Gamma)=(I-\Gamma\Stt)\Gamma$ by $L_\Gamma$ on the left and $R_\Gamma$ on the right. The second, $I+\Gamma R_\Gamma\Stt=L_\Gamma$, follows from the first: $I+L_\Gamma\Gamma\Stt=L_\Gamma[(I-\Gamma\Stt)+\Gamma\Stt]=L_\Gamma$. For two strict settings $X,Y$,
\begin{equation}
YR_Y-XR_X=L_YY-XR_X=L_Y\bigl[Y(I-\Stt X)-(I-Y\Stt)X\bigr]R_X=L_Y(Y-X)R_X,
\end{equation}
so
\begin{equation}
F(Y)-F(X)=\Set L_Y(Y-X)R_X\Ste,\qquad \dd F=\Set L_\Gamma(\dd\Gamma)R_\Gamma\Ste .
\label{eq:fd}
\end{equation}
The finite-difference formula \eqref{eq:fd} proves the rank inequality (9) of the main text: a product cannot have larger rank than any of its factors.

\subsection{Right energy balance}
For an external excitation $x$, the waves leaving the network toward the tuner are $b_{\mathrm t}=R_\Gamma\Ste x$ and those returning are $a_{\mathrm t}=\Gamma R_\Gamma\Ste x$. Collect the network inputs in $w=T_Rx$ with $T_R=\bigl[\begin{smallmatrix}I\\ \Gamma R_\Gamma\Ste\end{smallmatrix}\bigr]$; the network outputs are $Sw=\bigl[\begin{smallmatrix}Fx\\ R_\Gamma\Ste x\end{smallmatrix}\bigr]$. In words, the network receives the external wave $x$ and the wave $a_{\mathrm t}$ coming back from the tuner, and sends out $Fx$ and $b_{\mathrm t}$. Passivity of the network, $w^\dagger(I-S^\dagger S)w\ge0$, states that what it receives is at least what it sends out. Written out:
\begin{align}
w^\dagger(I-S^\dagger S)w&=\|x\|^2+\|\Gamma R_\Gamma\Ste x\|^2-\|Fx\|^2-\|R_\Gamma\Ste x\|^2\\
&=x^\dagger(I-F^\dagger F)x-(R_\Gamma\Ste x)^\dagger(I-\Gamma^\dagger\Gamma)(R_\Gamma\Ste x).
\end{align}
Since this holds for all $x$,
\begin{equation}
D_F^2=\Ste^\dagger R_\Gamma^\dagger D_\Gamma^2R_\Gamma\Ste+T_R^\dagger(I-S^\dagger S)T_R .
\label{eq:right}
\end{equation}
\emph{Physical reading:} the power lost by the whole device equals the power absorbed by the tuner plus the power absorbed by the network. Both terms are nonnegative. In particular $F^\dagger F\preceq I$: a passive network terminated by a passive tuner is passive.

\subsection{Left energy balance}
The adjoint network $S^\dagger$ terminated by $\Gamma^\dagger$ has response $F^\dagger$ (take the adjoint of every block relation). Applying \eqref{eq:right} to it gives
\begin{equation}
D_{F^\dagger}^2=\Set L_\Gamma D_{\Gamma^\dagger}^2L_\Gamma^\dagger\Set^\dagger+T_L^\dagger(I-SS^\dagger)T_L,\qquad T_L=\begin{bmatrix}I\\ \Gamma^\dagger L_\Gamma^\dagger\Set^\dagger\end{bmatrix}.
\label{eq:left}
\end{equation}

\subsection{Factorization}
Assume $F$ strict. Define
\begin{equation}
Q_\Gamma=D_\Gamma R_\Gamma\Ste D_F^{-1},\qquad P_\Gamma=D_{F^\dagger}^{-1}\Set L_\Gamma D_{\Gamma^\dagger}.
\end{equation}
Multiplying \eqref{eq:right} on both sides by $D_F^{-1}$ gives $I=Q_\Gamma^\dagger Q_\Gamma+D_F^{-1}T_R^\dagger(I-S^\dagger S)T_RD_F^{-1}$, hence $Q_\Gamma^\dagger Q_\Gamma\preceq I$ and $\|Q_\Gamma\|_2\le1$. In the same way \eqref{eq:left} gives $P_\Gamma P_\Gamma^\dagger\preceq I$. Now insert $D_{\Gamma^\dagger}D_{\Gamma^\dagger}^{-1}$ and $D_\Gamma^{-1}D_\Gamma$ into \eqref{eq:fd}:
\begin{equation}
D_{F^\dagger}^{-1}\,\dd F\,D_F^{-1}=\bigl(D_{F^\dagger}^{-1}\Set L_\Gamma D_{\Gamma^\dagger}\bigr)\bigl(D_{\Gamma^\dagger}^{-1}\dd\Gamma D_\Gamma^{-1}\bigr)\bigl(D_\Gamma R_\Gamma\Ste D_F^{-1}\bigr),
\end{equation}
which is $\mathcal V_F=P_\Gamma\mathcal V_\Gamma Q_\Gamma$. Taking norms, $\dd\Lambda_F=2\|\mathcal V_F\|_2\le2\|P_\Gamma\|_2\|\mathcal V_\Gamma\|_2\|Q_\Gamma\|_2\le\dd\Lambda_\Gamma$. Integrating along a trajectory proves $\Lambda_F\le\Lambda_\Gamma$; taking the infimum over trajectories joining two settings proves $d(F(X),F(Y))\le d(X,Y)$. The slack identities
\begin{equation}
I-Q_\Gamma^\dagger Q_\Gamma=D_F^{-1}T_R^\dagger(I-S^\dagger S)T_RD_F^{-1},\qquad
I-P_\Gamma P_\Gamma^\dagger=D_{F^\dagger}^{-1}T_L^\dagger(I-SS^\dagger)T_LD_{F^\dagger}^{-1}
\end{equation}
measure the dissipation for each normalized excitation. They do not force the operator norms below one whenever some excitation dissipates: $Q_\Gamma=\operatorname{diag}(1,a)$ with $0<a<1$ dissipates on one direction and still has norm one. Dissipation is also not the only way length is lost. In a lossless embedding $Q_\Gamma$ is an isometry and $P_\Gamma$ a coisometry, $Q_\Gamma^\dagger Q_\Gamma=I$ and $P_\Gamma P_\Gamma^\dagger=I$, and when the tuner has more ports than the observed response, $P_\Gamma\mathcal V_\Gamma Q_\Gamma$ can still be shorter than $\mathcal V_\Gamma$: a direction of tuner motion that the network does not reach, or does not return, drops out. A unitary network that couples the external port only to tuner port~1 gives $F(\Gamma)=\gamma_1$ for $\Gamma=\operatorname{diag}(\gamma_1,\gamma_2)$; moving $\gamma_2$ costs tuner length and moves nothing. Equality for a given motion therefore needs both: no dissipation on the waves that carry it, and full access to it and observation of it.

\subsection{Selected responses}
For unit vectors $u,v$, the maps $X\mapsto v^\dagger Xu$ and $X\mapsto Xu$ are themselves passive responses of the form $F$: they are the response of a network that routes a fixed input into $u$ and reads the output through $v$. Hence all statements hold for a single coefficient, for an output vector, and for any sub-block.

\subsection{Matrix distance}
The distance generated by the matrix metric is \cite{Harris1979,Helton1982}
\begin{equation}
d(X,Y)=2\atanh\bigl\|D_{X^\dagger}^{-1}(Y-X)(I-X^\dagger Y)^{-1}D_X\bigr\|_2 .
\label{eq:mdist}
\end{equation}
For scalars it reduces to $2\atanh\rho$. For two column vectors $x,y$ (one excitation, several outputs), writing $x=re$ and $y=pe+w$ with $\langle e,w\rangle=0$, the ball automorphism
$\varphi_x(y)=[(r-p)e-\sqrt{1-r^2}\,w]/(1-rp)$ sends $x$ to $0$, and the computation
\begin{equation}
1-\|\varphi_x(y)\|^2=\frac{|1-rp|^2-|r-p|^2-(1-r^2)\|w\|^2}{|1-rp|^2}=\frac{(1-\|x\|^2)(1-\|y\|^2)}{|1-\langle x,y\rangle|^2}
\end{equation}
gives the vector form of \eqref{eq:sech}:
\begin{equation}
\sech^2\frac{d(x,y)}2=\frac{(1-\|x\|^2)(1-\|y\|^2)}{|1-\langle x,y\rangle|^2}.
\label{eq:vecdist}
\end{equation}

\subsection{Equality and rank of full isometries}
For a scalar two-port, equality of input and output distances at one pair of distinct interior points forces the response to be an automorphism (Theorem~\ref{thm:SP}), and the slack identity then forces $S$ to be unitary. A map that preserves every distance of a $p\times q$ ball in a $p'\times q'$ ball through a terminated network must satisfy $p\le p'$ and $q\le q'$: setting $X=uv^\dagger$ in $\|P_\Gamma XQ_\Gamma\|_2=\|X\|_2$ gives $\|P_\Gamma u\|=\|Q_\Gamma^\dagger v\|=1$ for all unit $u,v$, so $P_\Gamma^\dagger P_\Gamma=I_p$ and $Q_\Gamma Q_\Gamma^\dagger=I_q$, which requires the stated dimensions \cite{LemmensWalsh2026}.

\section{Impedance Forms of the Metric}\label{sec:S5}
\emph{Scalar.} With $\gamma=(Z-Z_0)/(Z+Z_0)$:
\begin{equation}
\dd\gamma=\frac{2Z_0\,\dd Z}{(Z+Z_0)^2},\qquad 1-|\gamma|^2=\frac{|Z+Z_0|^2-|Z-Z_0|^2}{|Z+Z_0|^2}=\frac{4Z_0\Re Z}{|Z+Z_0|^2},
\end{equation}
so $2|\dd\gamma|/(1-|\gamma|^2)=|\dd Z|/\Re Z$.

\emph{Matrix.} Normalize the reference to $I$. Let $W=Z+I$, $\Gamma=(Z-I)W^{-1}=I-2W^{-1}$ and $R=\frac12(Z+Z^\dagger)\succ0$. Then $\dd\Gamma=-2\,\dd(W^{-1})=2W^{-1}\dd Z\,W^{-1}$. Expanding $\Gamma^\dagger\Gamma=(I-2W^{-\dagger})(I-2W^{-1})$,
\begin{equation}
I-\Gamma^\dagger\Gamma=2W^{-1}+2W^{-\dagger}-4W^{-\dagger}W^{-1}=W^{-\dagger}\bigl(2W+2W^\dagger-4I\bigr)W^{-1}=4W^{-\dagger}RW^{-1},
\end{equation}
since $2W+2W^\dagger-4I=2(Z+Z^\dagger)=4R$; in the same way $I-\Gamma\Gamma^\dagger=4W^{-1}RW^{-\dagger}$.
The matrices $U_L=2D_{\Gamma^\dagger}^{-1}W^{-1}R^{1/2}$ and $U_R=2R^{1/2}W^{-1}D_\Gamma^{-1}$ satisfy $U_LU_L^\dagger=D_{\Gamma^\dagger}^{-1}\bigl(4W^{-1}RW^{-\dagger}\bigr)D_{\Gamma^\dagger}^{-1}=D_{\Gamma^\dagger}^{-1}D_{\Gamma^\dagger}^2D_{\Gamma^\dagger}^{-1}=I$ and, in the same way, $U_R^\dagger U_R=I$; being square, they are unitary. Therefore
\begin{equation}
\mathcal V_\Gamma=D_{\Gamma^\dagger}^{-1}\dd\Gamma D_\Gamma^{-1}=\tfrac12U_L\bigl[R^{-1/2}\dd Z\,R^{-1/2}\bigr]U_R,\qquad
\dd\Lambda=\bigl\|R^{-1/2}\dd Z\,R^{-1/2}\bigr\|_2,
\end{equation}
without assuming that $R$ and $Z$ commute. The same form holds for admittance matrices with $G=\frac12(Y+Y^\dagger)$.

\section{Tuning Budgets of Elements}\label{sec:S6}
\subsection{Constant-resistance capacitor (varactor, MEMS)}
$Z=R_s-j/(\omega C)$, so $|\dd Z|=\dd C/(\omega C^2)$ and
\begin{equation}
\Lambda=\int_{C_{\min}}^{C_{\max}}\frac{\dd C}{\omega R_sC^2}=\frac1{\omega R_s}\Bigl(\frac1{C_{\min}}-\frac1{C_{\max}}\Bigr)=Q_{\max}\Bigl(1-\frac1n\Bigr)=Q_{\max}-Q_{\min},\qquad Q_{\max}=\frac1{\omega R_sC_{\min}} .
\end{equation}
Here $Q(C)=1/(\omega R_sC)$ is largest at $C_{\min}$, so the budget is the range of the varactor's quality factor over the sweep. The end states differ only in reactance, $\kappa=\Delta X/R_s=\Lambda$, $d=2\asinh(\Lambda/2)$ and $K=e^d$. The line $\Re Z=R_s$ is a horocycle (Note~S2), along which the distance between two points grows monotonically with their separation, so the end states are the most distant pair. A bias-dependent $R_s(V)$ is handled by the general integral below.

\subsection{\texorpdfstring{Constant-$Q$ capacitor or inductor}{Constant-Q capacitor or inductor}}
A tunable dielectric capacitor with loss tangent $\tan\delta=1/Q$ has $Z=1/[j\omega C(1-j\tan\delta)]=c/C$, $c=1/[\omega(\tan\delta+j)]$. Then $|\dd Z|/\Re Z=(|c|/\Re c)\,\dd C/C$ and $|c|/\Re c=\sqrt{1+\tan^2\delta}/\tan\delta=\sqrt{1+Q^2}$, so
\begin{equation}
\Lambda=\sqrt{1+Q^2}\,\ln n,\qquad \kappa^2=\frac{|c|^2}{(\Re c)^2}\frac{(C_2-C_1)^2}{C_1C_2}=(1+Q^2)\frac{(n-1)^2}{n}.
\end{equation}
The trajectory is a ray from the origin of the impedance plane, a hypercycle at fixed distance from the real-axis geodesic. A tunable inductor (ferrite, magnetic film) with constant $Q$ has the same budget.

\emph{Bias-dependent loss tangent.} If $t=\tan\delta$ changes along the sweep, both the capacitance and the dissipation move $Z$. In admittance form $Y=\omega Ct+j\omega C$, so $\dd G=\omega(t\,\dd C+C\,\dd t)$, $\dd B=\omega\,\dd C$ and
\begin{equation}
\dd\Lambda=\frac{|\dd Y|}{\Re Y}=\frac{\dd C}{C}\sqrt{Q^2+\Bigl(1+\frac{\dd\ln t}{\dd\ln C}\Bigr)^2},\qquad Q=\frac1t .
\label{eq:varQ}
\end{equation}
For constant $t$ this is $\sqrt{1+Q^2}\,\dd C/C$. Since $Q\gg1$ in practice, the extra term is small unless the loss tangent changes steeply, but an exclusion test must use \eqref{eq:varQ}, or the measured trajectory, rather than the constant-$Q$ form. A numerical check is in Note~S21.

\subsection{Switches}
For $Z_{\mathrm{on}}=R_{\mathrm{on}}$ and $Z_{\mathrm{off}}=R_{\mathrm{off}}-j/(\omega C_{\mathrm{off}})$, Note~S3 gives
\begin{equation}
\kappa^2=\frac{(R_{\mathrm{off}}-R_{\mathrm{on}})^2+1/(\omega C_{\mathrm{off}})^2}{R_{\mathrm{on}}R_{\mathrm{off}}}=\Bigl(\frac{f_c}f\Bigr)^2+\frac{(R_{\mathrm{off}}-R_{\mathrm{on}})^2}{R_{\mathrm{on}}R_{\mathrm{off}}},\qquad f_c=\frac{1}{2\pi C_{\mathrm{off}}\sqrt{R_{\mathrm{on}}R_{\mathrm{off}}}} .
\end{equation}
The first term dominates while $1/(\omega C_{\mathrm{off}})\gg|R_{\mathrm{off}}-R_{\mathrm{on}}|$; only in that reactive-dominated regime is $\kappa\approx f_c/f$ and $K\approx(f_c/f)^2$. The second term does not depend on frequency and sets a floor on $\kappa$ that frequency-ceiling calculations must keep. When a datasheet gives only $R_{\mathrm{on}}$ and $C_{\mathrm{off}}$, the familiar $f_c=1/(2\pi R_{\mathrm{on}}C_{\mathrm{off}})$ corresponds to $R_{\mathrm{off}}=R_{\mathrm{on}}$; an off state with no resistance lies on the lossless boundary and needs a measured value. Fixed series inductance belongs to the network and leaves $\kappa$ and $K$ unchanged.

\subsection{Packages, parasitics and fixed loss}
A fixed lossless parasitic (bond-wire or package inductance, pad capacitance) is part of the network: it moves the trajectory by an automorphism and changes nothing. A fixed lossy element in series, $R_p$, may be left in the network or absorbed into the tuner. Absorbing it gives the valid and tighter budget $Q_{\max}(1-1/n)R_s/(R_s+R_p)$ for a constant-$R_s$ element, because the absorbed trajectory is $\Re Z=R_s+R_p$ with the same reactance range.

\subsection{Stacking, joint drive and independent elements}
Series or parallel stacks of $k$ identical elements scale $Z$ by $k$ or $1/k$, leaving $|\dd Z|/\Re Z$ unchanged. Identical elements driven together form $\Gamma=\gamma I_k$; its defects are $\sqrt{1-|\gamma|^2}I$ and $\|\mathcal V_\Gamma\|_2=|\dd\gamma|/(1-|\gamma|^2)$, the single-element value. For independent elements $\Gamma=\bigoplus_k\Gamma_k$ the defects and $\mathcal V_\Gamma$ are block diagonal, so $\dd\Lambda=\max_k\dd\Lambda_k$; \eqref{eq:mdist} is block diagonal too and $d=\max_kd_k$.

\subsection{Distributed tuners and tunable dielectrics}
A loaded line (varactor-loaded, MEMS-loaded) is a cascade of identical cells whose tunable elements are driven together. A tunable dielectric filling a structure can be discretized into small cells; in the quasi-static limit of a fine mesh each cell is a capacitor whose value is proportional to the local permittivity, so all tunable cells are scaled copies $Z_k=c_kZ_{\mathrm{cell}}$ of one impedance law. Because $|\dd Z|/\Re Z$ is scale invariant, every cell has the same line element at every bias, and the block-diagonal tuner has the budget of one cell. For permittivity $\varepsilon_{\mathrm r}$ with loss tangent $1/Q(\varepsilon_{\mathrm r})$ along the field,
\begin{equation}
\Lambda=\int_{\varepsilon_{\mathrm r,\min}}^{\varepsilon_{\mathrm r,\max}}\sqrt{Q^2+\Bigl(1+\frac{\dd\ln\tan\delta}{\dd\ln\varepsilon_{\mathrm r}}\Bigr)^2}\,\frac{\dd\varepsilon_{\mathrm r}}{\varepsilon_{\mathrm r}},
\end{equation}
by \eqref{eq:varQ} with $C\propto\varepsilon_{\mathrm r}$. For constant loss tangent, $\Lambda=\sqrt{1+Q^2}\ln[1/(1-\tau)]$ with $\tau=1-\varepsilon_{\mathrm r,\min}/\varepsilon_{\mathrm r,\max}$; for small $\tau$, $\ln[1/(1-\tau)]\approx\tau$ and $\Lambda\approx\tau/\tan\delta$, the material figure of merit used for liquid crystals \cite{Jakoby2020}. Example: $\tau=0.24$, $\tan\delta=0.0143$ gives $\Lambda=19.2$ (constant-$Q$ approximation), against $\tau/\tan\delta=16.8$. Conductor and fixed-dielectric losses belong to the network and only reduce the delivered length. The assumption is that the material acts through its local capacitance along a fixed field direction, so that the cells remain scaled copies of one impedance law. In an anisotropic medium whose principal axes rotate relative to the field (liquid crystal in a non-aligned geometry, for example), the cell couples field components and becomes a multiport; its budget is then the matrix integral of Note~S5 over the cell admittance tensor, and the scalar effective-permittivity form is not sufficient.

\subsection{Budgets from measured data}
Measure the calibrated $S$-matrix of the tuner block at its own reference planes for a set of bias points $V_0,\dots,V_N$ (de-embed fixed fixtures; they are part of the network anyway). Then
\begin{enumerate}
\item for two states, compute $d$ from \eqref{eq:mdist} (scalar: $2\atanh\rho$);
\item for a sweep, sum $\Lambda\approx\sum_i d(\Gamma(V_i),\Gamma(V_{i+1}))$; for exact samples, refining the bias grid increases the sum toward the path length, so a coarse grid gives a lower estimate. Measurement noise can lengthen the sampled polygon, and motion between samples is not seen, so a sampled sum cannot by itself exclude a design; an exclusion test needs an upper bound from a model or from monotonicity between samples, as in E3;
\item if the element is described by an impedance law, integrate \eqref{eq:halfplane} directly.
\end{enumerate}

\subsection{Frequency scaling}
With constant parameters, $\Lambda_{\mathrm{var}}=(1-1/n)f_{cv}/f$ with $f_{cv}=1/(2\pi R_sC_{\min})$; in the reactive-dominated regime the switching ratio $\kappa$ scales as $f_c/f$ and the CQF as $(f_c/f)^2$, and the full expression of Note~S6.3 applies otherwise; the constant-$Q$ budget follows $Q(f)$.

\section{Phase Shifters, Figure of Merit and True-Time Delay}\label{sec:S7}
\subsection{Line element}
For a nonzero selected coefficient $t=e^{-\ell+j\phi}$, $\dd t=t(-\dd\ell+j\dd\phi)$ and $1-|t|^2=1-e^{-2\ell}$. Hence
\begin{equation}
\dd\Lambda_t=\frac{2e^{-\ell}\sqrt{\dd\ell^2+\dd\phi^2}}{1-e^{-2\ell}}=\frac{\sqrt{\dd\ell^2+\dd\phi^2}}{\sinh\ell} .
\end{equation}

\subsection{Peak, mean and occupancy bounds}
Let $\Phi=\int|\dd\phi|$ be the phase travel, the total variation of the unwrapped phase. The phase range $\max\phi-\min\phi$ never exceeds $\Phi$, and the two are equal when the phase is monotone; a trajectory whose phase moves back and forth gains travel without gaining range. Every bound below holds for the range as well. Since $\sqrt{\dd\ell^2+\dd\phi^2}\ge|\dd\phi|$,
\begin{equation}
\Lambda\ge L[t]\ge\int\frac{|\dd\phi|}{\sinh\ell}.
\label{eq:budget}
\end{equation}
\emph{Peak:} $\sinh\ell\le\sinh\ell_{\max}$ gives $\Lambda\ge\Phi/\sinh\ell_{\max}$.
\emph{Mean:} with the probability measure $\dd\mu=|\dd\phi|/\Phi$ and the convex function $\csch\ell$ (its second derivative is $\csch\ell\,(1+2\csch^2\ell)>0$), Jensen's inequality gives $\Lambda/\Phi\ge\int\csch\ell\,\dd\mu\ge\csch\bar\ell_\phi$ with $\bar\ell_\phi=\int\ell\,\dd\mu$, i.e.\ $\bar\ell_\phi\ge\asinh(\Phi/\Lambda)$.
\emph{Occupancy:} restricting \eqref{eq:budget} to the part of the path with $\ell\le\ell_0$ gives $\Phi_{\ell\le\ell_0}\le\Lambda\sinh\ell_0$.
\emph{Equality} requires constant loss, no loss variation and lossless transfer. A circulator routing the input to the element and back gives $F=\Gamma$; with $\Gamma=\rho e^{j\theta}$, $0\le\theta\le2\pi$, $\Lambda=4\pi\rho/(1-\rho^2)=2\pi/\sinh(-\ln\rho)$ and all bounds hold with equality. This defines the full-turn price $\Lambda_{360}(\rho)=4\pi\rho/(1-\rho^2)$ printed on the price rings.

\subsection{Figure of merit}
$\mathrm{FoM}=\Phi^\circ/\IL_{\max}=(180/\pi)\Phi\,/\,[(20/\ln10)\ell_{\max}]=6.596\,\Phi/\ell_{\max}$. Inserting $\Phi\le\Lambda\sinh\ell_{\max}$ gives $\mathrm{FoM}\le6.596\Lambda\sinh\ell_{\max}/\ell_{\max}$. Since $\sinh x/x=1+x^2/6+\dots$, the factor is $1.005$ at $1.56$\,dB and $1.24$ at $10$\,dB. The exact inequality is the universal bound; $\mathrm{FoM}\le6.6\Lambda$ is its low-loss form.

\subsection{Varactor ceiling}
With $\Lambda=Q_{\max}(1-1/n)$, the peak bound gives $\ell_{\max}\ge\asinh[\Phi/(Q_{\max}(1-1/n))]$. Writing $Q_{\max}=f_{cv}/f$ and solving for $f$ gives $f\le f_{cv}(1-1/n)\sinh\ell_{\max}/\Phi$.

\subsection{True-time delay}
A delay change $\Delta\tau$ is a phase change $2\pi f\Delta\tau$ at frequency $f$; the largest phase travel occurs at the top of the band, $f_H$. The peak bound at $f_H$ with $\Lambda(f_H)=f_{cv}(1-1/n)/f_H$ gives
\begin{equation}
2\pi f_H\Delta\tau\le\frac{f_{cv}(1-1/n)}{f_H}\sinh\ell_{\max}\quad\Longrightarrow\quad\Delta\tau f_H^2\le\frac{f_{cv}(1-1/n)\sinh\ell_{\max}}{2\pi}.
\end{equation}
For $f_{cv}=350$\,GHz, $n=6$, $1$\,dB ($\ell_{\max}=0.1151$) and $f_H=10$\,GHz the right side gives $\Delta\tau\le53.6$\,ps.

\subsection{Winding}
A closed trajectory with winding number $w$ about the origin has $\Phi\ge2\pi|w|$, so $\bar\ell_\phi,\ell_{\max}\ge\asinh(2\pi|w|/\Lambda)$. Without a finite budget winding alone gives no loss floor: circles $\rho e^{j\theta}$ with $\rho\to1$ have diverging length.

\subsection{Noise figure}
A passive two-port at ambient temperature $T_0$ has noise factor $F_n=1/G_{\mathrm{av}}$ \cite{Pozar2012}. With a matched source, $G_{\mathrm{av}}=|S_{21}|^2/(1-|S_{22}|^2)$, which equals $|S_{21}|^2$ for a matched output. Then $\mathrm{NF}=\IL$ and every loss bound of this note is a noise-figure bound.

\section{Element Count}\label{sec:S8}
Let the $N$ elements have reflections $\gamma_k(C)=(\alpha_kC+\beta_k)/(\mu_kC+\nu_k)$ in a common real control $C$ (identical varactors in $C$; constant-$R_s$ and constant-$Q$ laws are of this form), $\Gamma=\operatorname{diag}(\gamma_k)$, and let $t=v^\dagger F(\Gamma)u$.

\emph{Step 1: $t$ is a ratio of multi-affine functions.} With $M=I-\Stt\Gamma$, $M^{-1}=\adj M/\det M$, so
\begin{equation}
t=\frac{v^\dagger\See u\,\det M+v^\dagger\Set\Gamma\adj(M)\Ste u}{\det M}.
\end{equation}
Column $k$ of $M$ depends only on $\gamma_k$ and is affine in it, so $\det M$ is affine in each $\gamma_k$ separately. The entry $(\adj M)_{kl}$ is a cofactor that omits column $k$ of $M$; multiplied by $\gamma_k$ it is again affine in each $\gamma_i$. The numerator and denominator are multi-affine in $(\gamma_1,\dots,\gamma_N)$.

\emph{Step 2: polynomials in $C$.} Multiplying numerator and denominator by $\prod_k(\mu_kC+\nu_k)$ turns each into a polynomial of degree at most $N$ in $C$; the common factor is nonzero on the sweep because the $\gamma_k$ are finite.

\emph{Step 3: argument of each factor.} Write a polynomial as $a\prod_i(C-z_i)$. For real $C$ on an interval that contains no root, $\frac{\dd}{\dd C}\arg(C-z_i)=\Im z_i/|C-z_i|^2$ has constant sign, and the argument changes by less than $\pi$ (it would need $C$ to run over the whole real line to reach $\pi$). A real root outside the interval contributes a constant argument.

\emph{Step 4: total.} $\arg t$ is the sum of at most $N$ numerator arguments minus at most $N$ denominator arguments, each monotone with variation below $\pi$. Its total variation is therefore below $2N\pi$, i.e.\ $\Phi<N\cdot360^\circ$.

The bound concerns the structure of the response; a specific topology may use fewer degrees of freedom. In a reflection-type phase shifter with identical arms and an ideal hybrid, $S_{21}$ is proportional to the arm reflection, which has degree equal to the number of elements in one arm.

\section{The Horocycle Law of Single-Element Cells}\label{sec:S9}
\subsection{Statement and derivation}
\emph{Scope.} The law requires one propagating external channel (a single Floquet mode in a single polarization, no grating lobes, no cross-polarized reflection), a lossless fixed structure and a constant-$R_s$ element. A cell with transmission, cross-polarization, diffraction orders or loss has a multiport or lossy embedding; its trajectory is then a passive image of the horocycle, generally not a horocycle, with length at most $\Lambda$, and Rule~1 bounds it.

\emph{Step 1: the cell is an automorphism.} A reflecting cell with one tuning element, one propagating Floquet mode and lossless fixed structure is a lossless two-port (Floquet port, element port) terminated by the element. By Lemma~\ref{lem:lossless} its reflection is $\Gamma_{\mathrm{cell}}=\varphi(\gamma)$ with $\varphi$ an automorphism, so the cell trajectory has exactly the length of the element trajectory.

\emph{Step 2: the element moves on a horocycle.} A constant-$R_s$ element satisfies $\Re Z=R_s$, a constant-resistance circle of the Smith chart, which is a horocycle (Note~S2). Automorphisms map circles to circles and the unit circle to itself, so they map horocycles to horocycles. The cell reflection lies on a horocycle segment of length $\Lambda=Q_{\max}(1-1/n)$.

\emph{Step 3: arc length on a horocycle.} Rotate so the tangent point is $1$. A horocycle of Euclidean radius $s$ is $z(\vartheta)=1-s+se^{j\vartheta}$. Then
\begin{equation}
|z|^2=(1-s)^2+s^2+2s(1-s)\cos\vartheta,\qquad 1-|z|^2=2s(1-s)(1-\cos\vartheta)=4s(1-s)\sin^2\tfrac\vartheta2,
\end{equation}
and $|\dd z|=s\,\dd\vartheta$, so
\begin{equation}
\dd\Lambda=\frac{2s\,\dd\vartheta}{4s(1-s)\sin^2(\vartheta/2)}=\frac{\dd\vartheta}{2(1-s)\sin^2(\vartheta/2)},\qquad
\Lambda(\vartheta_a,\vartheta_b)=\frac{\cot(\vartheta_a/2)-\cot(\vartheta_b/2)}{1-s}.
\label{eq:horolen}
\end{equation}
The tangent point $\vartheta=0$ is at infinite distance, as it lies on the lossless boundary.

\emph{Step 4: amplitude and phase.} $|z|$ is largest ($\to1$) near the tangent point and smallest at $\vartheta=\pi$, where $|z|=|1-2s|$. The horocycle encloses the origin when $s>\frac12$; only then can the phase range exceed $180^\circ$. Since $\dd\arg z/\dd\vartheta=s[s+(1-s)\cos\vartheta]/|z|^2$ and $s+(1-s)\cos\vartheta\ge2s-1$, the phase is monotone along the whole horocycle when $s>\frac12$; when $s<\frac12$ it has turning points, although short segments can still have monotone phase. For $s>\frac12$, a segment symmetric about $\vartheta=\pi$, $\vartheta\in[\vartheta_1,2\pi-\vartheta_1]$, has length $2\cot(\vartheta_1/2)/(1-s)$, minimum amplitude $2s-1$, and phase travel $\Phi=2\pi-2\arg(1-s+se^{j\vartheta_1})$.

\emph{Step 5: every segment of a given length is realizable.} Automorphisms act transitively on horocycles, and the automorphisms that preserve a given horocycle slide it along itself, preserving arc length. Hence any two horocycle segments of equal length are images of each other under an automorphism, and by Lemma~\ref{lem:lossless} a lossless (line--transformer--line) cell maps the element's segment to any chosen segment of the same length.

\subsection{Consequences}
The amplitude--phase coupling of single-varactor cells is a geometric necessity of lossless cells: a horocycle that wraps the origin must pass within $2s-1$ of it. The designer chooses $s$ and the placement; the varactor fixes the length. For a given $\Lambda$ and $\Phi$, solving \eqref{eq:horolen} for $s$ with the symmetric placement gives an attainable minimum amplitude. Rule~1 bounds every passive cell, lossless or lossy, by the constant-amplitude ceiling $e^{-\asinh(\Phi/\Lambda)}$. The worked example of the main text is in Note~S18.

\section{Two-State Devices and Their Attainment}\label{sec:S10}
\subsection{The master inequality}
For two element states of CQF $K$ (switching ratio $\kappa$, Note~S3) and two selected responses $t_1,t_2$, Theorem~1 gives $d(t_1,t_2)\le d(\gamma_1,\gamma_2)$, and Note~S3 turns this into
\begin{equation}
|t_1-t_2|^2\le\frac{\kappa^2}4(1-|t_1|^2)(1-|t_2|^2).
\label{eq:master}
\end{equation}

\subsection{SPST switches and attenuator steps}
If the phases are free, the left side is smallest when the phases are equal: $|t_1-t_2|\ge\bigl||t_1|-|t_2|\bigr|$. With $|t_i|=L_i^{-1/2}$:
\begin{equation}
\frac{(|t_1|-|t_2|)^2}{(1-|t_1|^2)(1-|t_2|^2)}=\frac{(L_1^{-1/2}-L_2^{-1/2})^2}{(1-L_1^{-1})(1-L_2^{-1})}=\frac{(\sqrt{L_2}-\sqrt{L_1})^2}{(L_1-1)(L_2-1)}\le\frac{\kappa^2}4 .
\end{equation}
With perfect isolation, $t_2=0$: $|t_1|^2/(1-|t_1|^2)\le\kappa^2/4=\sinh^2(d/2)$, i.e.\ $|t_1|\le\tanh(d/2)=r_{\mathrm c}=(K-1)/(K+1)$ and $\IL_{\mathrm{on}}\ge20\log_{10}[(K+1)/(K-1)]=10\log_{10}(1+4/\kappa^2)$\,dB.

\subsection{Phase bits}
For equal amplitudes $|t_1|=|t_2|=a=e^{-\ell}$ and phase step $\Delta\phi$: $|t_1-t_2|^2=4a^2\sin^2(\Delta\phi/2)$, and with $a=e^{-\ell}$, $(1-a^2)/(2a)=(e^{\ell}-e^{-\ell})/2=\sinh\ell$. Dividing \eqref{eq:master} by $(1-a^2)^2$ therefore gives
\begin{equation}
\frac{\sin^2(\Delta\phi/2)}{\sinh^2\ell}\le\frac{\kappa^2}4\quad\Longleftrightarrow\quad\sinh\ell\ge\frac{2\sin(\Delta\phi/2)}{\kappa}=\frac{2\sqrt K\sin(\Delta\phi/2)}{K-1}.
\end{equation}
For $\Delta\phi=\pi$ the two responses $\pm a$ are $4\atanh a$ apart, so $a\le\tanh(d/4)=r_{\mathrm c}/(1+\sqrt{1-r_{\mathrm c}^2})$ (half-angle formula for $\tanh$). With $K=e^d$, $\tanh(d/4)=(\sqrt K-1)/(\sqrt K+1)$, which is eq.~(9) of Vendik \emph{et al.} \cite{Vendik2000}, derived there for a reflection-type bit with real transformed resistances; here it bounds every passive network.

\emph{Small-loss forms.} For $\kappa\gg1$, $\ell\approx2\sin(\Delta\phi/2)/\kappa$, so $|t|\approx1-2\sin(\Delta\phi/2)/\sqrt K$, eq.~(11) of \cite{Vendik2000}, and the dissipated fraction is $1-e^{-2\ell}\approx4(f/f_c)\sin(\Delta\phi/2)$, the perturbation result of Hines \cite{Hines1964} as quoted by White \cite{White1974}. In decibels, a $180^\circ$ bit loses at least $17.4f/f_c$\,dB; White tabulates about $18f/f_c$\,dB for one-bit diode shifters.

\subsection{Attainment by a reflection-type circuit}
Let an ideal $90^\circ$ hybrid have its input and output on an isolated port pair and two identical reflective arms on the other pair. With identical arm reflections $\Gamma_a$ the input is matched and $S_{21}=e^{j\theta_{\mathrm h}}\Gamma_a$, with $\theta_{\mathrm h}$ fixed by the hybrid convention. Each arm is a lossless reciprocal two-port terminating the element, so $\Gamma_a=\varphi(\gamma)$ for an automorphism of our choice (Lemma~\ref{lem:lossless}). Automorphisms are transitive on pairs at equal distance: map $\gamma_1$ to $0$, rotate so that $\gamma_2$ lands on the ray of the target pair, and map $0$ to the first target. Hence any target pair $(t_1,t_2)$ with $d(t_1,t_2)=d(\gamma_1,\gamma_2)$ is realized, and \eqref{eq:master} holds with equality. The two identical arms switch together, so their joint budget equals that of one element, although the circuit uses two physical elements. The construction attains a specified pair of states at one frequency; a filter meeting a whole passband and stopband specification is a separate synthesis problem. A numerical construction (Note~S21) reproduces the $180^\circ$ bit and the SPST at their bounds with $S_{11}=0$.

\section{Routing Switches and Polarization Switching}\label{sec:S11}
\subsection{Exact minimum separation for two output states}
Let one fixed input produce, in two states, outputs $x=r\psi_1$ and $y=se^{j\alpha}\psi_2$ with unit $\psi_i$, overlap $c=|\langle\psi_1,\psi_2\rangle|<1$, and efficiencies $r^2,s^2\ge\eta$. By \eqref{eq:vecdist}, $\sech^2(\delta/2)=(1-r^2)(1-s^2)/|1-rse^{j\alpha}\langle\psi_1,\psi_2\rangle|^2$. The phase $\alpha$ is free, and the denominator is smallest when $rse^{j\alpha}\langle\psi_1,\psi_2\rangle$ is real and positive, where it equals $(1-crs)^2$. So the smallest distance belongs to the largest value of
\begin{equation}
G(r,s)=\frac{\sqrt{(1-r^2)(1-s^2)}}{1-crs}
\end{equation}
after choosing $\alpha$ to align the overlap phase. An interior stationary point would need $r=cs$ and $s=cr$, impossible for $c<1$. On the edge $r=\sqrt\eta$, $\partial\log G/\partial s=(c\sqrt\eta-s)/[(1-s^2)(1-c\sqrt\eta s)]<0$ for $s\ge\sqrt\eta$; the other edge is symmetric. So the maximum is at $r=s=\sqrt\eta$, $\cosh(\delta/2)=(1-\eta c)/(1-\eta)$, and
\begin{equation}
\delta_{\eta,c}=2\acosh\frac{1-\eta c}{1-\eta},\qquad \eta\le\frac{1-\sech(\mathcal D/2)}{1-c\,\sech(\mathcal D/2)}\ \text{ for settings }\mathcal D\text{ apart.}
\end{equation}

\subsection{\texorpdfstring{SPDT, SP$n$T and T/R switches}{SPDT, SPnT and T/R switches}}
A routing switch toggles one switch or complementary sets of switches; the tuner states are $\operatorname{diag}(\gamma_{\mathrm{on}},\gamma_{\mathrm{off}},\dots)$ and permutations. By Note~S6 their distance is the single-switch distance $d_{\mathrm{sw}}$ with $\cosh(d_{\mathrm{sw}}/2)=\sqrt{1+\kappa^2/4}=(K+1)/(2\sqrt K)$. Requiring $\delta_{\eta,0}\le d_{\mathrm{sw}}$:
\begin{equation}
\frac1{1-\eta}\le\frac{K+1}{2\sqrt K}\quad\Longrightarrow\quad1-\eta\ge\frac{2\sqrt K}{K+1}=\sqrt{1-r_{\mathrm c}^2}\approx\frac2{\kappa}.
\end{equation}
For SP$n$T every pair of throws obeys the same inequality, so the floor is the same.

\subsection{Comparison with the quarter-wave shunt SPDT}
In the classical shunt SPDT \cite{Pozar2012,White1974}, each arm has a shunt switch a quarter wavelength from the junction. In the isolated arm the on-switch ($R_{\mathrm{on}}$) is transformed to a shunt conductance $R_{\mathrm{on}}/Z^2$ at the junction; in the through arm the off-switch presents the shunt conductance $\Re[1/(R_{\mathrm{off}}+1/(j\omega C_{\mathrm{off}}))]\approx\omega^2C_{\mathrm{off}}^2R_{\mathrm{off}}$ (its susceptance is tuned out losslessly). A small shunt conductance $G$ on a matched line of impedance $Z$ absorbs the fraction $GZ$, so the loss fraction is
\begin{equation}
1-\eta\approx\frac{R_{\mathrm{on}}}Z+\omega^2C_{\mathrm{off}}^2R_{\mathrm{off}}Z\ \ge\ 2\omega C_{\mathrm{off}}\sqrt{R_{\mathrm{on}}R_{\mathrm{off}}}=\frac{2f}{f_c},
\end{equation}
with the minimum at $Z=1/(\omega C_{\mathrm{off}})\sqrt{R_{\mathrm{on}}/R_{\mathrm{off}}}$, reached through a lossless impedance transformation. To leading order this equals the bound $2/\kappa\approx2f/f_c$.

\subsection{Polarization switching}
Two radiated patterns of the same shape and opposite pure circular polarization are orthogonal ($c=0$). At $\eta=0.9$, $\delta=2\acosh(10)=5.99$, so $K=e^\delta=398$ ($\kappa^2=396$). Linear versus circular polarization of the same shape has $c=1/\sqrt2$ and $\delta=2\acosh[(1-0.9/\sqrt2)/0.1]=3.93$.

\section{Attenuators, Vector Modulators and Impedance Tuners}\label{sec:S12}
\subsection{Variable attenuators}
Let $\ell_2\ge\ell_1>0$. With constant phase, $\dd\Lambda=\dd\ell/\sinh\ell$ and $\int\dd\ell/\sinh\ell=\ln\tanh(\ell/2)$, so $\Lambda\ge\ln[\tanh(\ell_2/2)/\tanh(\ell_1/2)]$. The radius is a geodesic, so this is also the distance between end states of equal phase and is attained by transferring a radial element trajectory without loss. Equal phase gives the smallest distance between the two loss levels; end states of different phase are farther apart, and a path whose phase varies in between is longer than the radial one. In every case the bound holds. For $0.5$--$30$\,dB, $\ell_1=0.0576$, $\ell_2=3.454$ and $\Lambda\ge3.49$. Digital attenuator steps use Note~S10.

\subsection{Covering argument}
A set of responses must contain a target disk of hyperbolic radius $D$, and every target must be within $\varepsilon$ of an achieved response. Balls of radius $\varepsilon$ have area $4\pi\sinh^2(\varepsilon/2)$ wherever they are, so at least $\sinh^2(D/2)/\sinh^2(\varepsilon/2)$ of them are needed.

\emph{Switched tuners.} For VSWR $S$, $D=\ln S$ and $\sinh^2(\ln S/2)=(\sqrt S-1/\sqrt S)^2/4=(S-1)^2/(4S)$. A mismatch loss $\mathrm{ML}$ dB corresponds to $\cosh^2(\varepsilon/2)=10^{\mathrm{ML}/10}$, i.e.\ $\sinh^2(\varepsilon/2)=10^{\mathrm{ML}/10}-1$. Therefore $M\ge(S-1)^2/[4S(10^{\mathrm{ML}/10}-1)]$. VSWR $10$ at $0.1$\,dB gives $86.9$, so $87$ states; VSWR $3$ at $0.5$\,dB gives $2.73$, so $3$ states.

\emph{Continuous tuners, exact coverage.} Suppose the achieved responses contain the whole target disk. A control path of length $\Lambda_k$ is covered by $\lceil\Lambda_k/(2\varepsilon)\rceil$ intervals of radius $\varepsilon$. For two independent elements the products of these intervals cover the control set in the maximum metric, and by the distance law their images lie in balls of radius $\varepsilon$ that cover the response set, and with it the target. Hence $\lceil\Lambda_1/2\varepsilon\rceil\lceil\Lambda_2/2\varepsilon\rceil\ge\sinh^2(D/2)/\sinh^2(\varepsilon/2)$ for every $\varepsilon>0$; letting $\varepsilon\to0$ gives $\Lambda_1\Lambda_2\ge16\sinh^2(D/2)$: $4(S-1)^2/S$ for VSWR coverage ($32.4$ at $S=10$) and $16\rho^2/(1-\rho^2)$ for a complex-weight disk $|t|\le\rho$ ($28.4$ at $\rho=0.8$). A single path of finite length is covered by $O(1/\varepsilon)$ balls and cannot contain any area.

\emph{Continuous tuners, finite tolerance.} If every target need only lie within $\varepsilon_0$ of an achieved response, the same covering, now with balls of radius $\varepsilon_0+\varepsilon$ and $N_k=\max(1,\lceil\Lambda_k/2\varepsilon\rceil)$ intervals per path, gives
\begin{equation}
N_1N_2\ \ge\ \frac{\sinh^2(D/2)}{\sinh^2[(\varepsilon_0+\varepsilon)/2]}\qquad\text{for every }\varepsilon>0 .
\end{equation}
This requirement is weaker than exact coverage and vanishes when one fixed response already lies within $\varepsilon_0$ of the whole disk: for loads within VSWR $3$, a fixed match has worst-case mismatch efficiency $4S/(S+1)^2=0.75$, about $1.25$\,dB, so a $2$\,dB allowance needs no tuning. One continuous element can meet a finite tolerance over an area, because the neighborhood of a curve has area.

\subsection{Reach}
From a starting presented impedance $Z_A$, all presented impedances satisfy $d(Z,Z_A)\le\Lambda$. With $r=\tanh(\Lambda/2)$, $|Z-Z_A|^2\le r^2|Z+Z_A^*|^2$; using $(1+r^2)/(1-r^2)=\cosh\Lambda$ and $2r/(1-r^2)=\sinh\Lambda$ this becomes $(R-R_A\cosh\Lambda)^2+(X-X_A)^2\le R_A^2\sinh^2\Lambda$, and $R_Ae^{-\Lambda}\le R\le R_Ae^{\Lambda}$.

\section{Tunable and Switched Filters}\label{sec:S13}
At any fixed frequency $f_0$, a filter is a two-port whose transmission $t(f_0)$ is a passive response of its tuning elements; Theorem~1 applies at that frequency.
\emph{Two-state tuning.} Moving $f_0$ from passband loss $L_{\mathrm p}$ to rejection $L_{\mathrm s}$ requires $\kappa(f_0)^2\ge4(\sqrt{L_{\mathrm s}}-\sqrt{L_{\mathrm p}})^2/[(L_{\mathrm p}-1)(L_{\mathrm s}-1)]$. For $1$\,dB and $30$\,dB, $L_{\mathrm p}=1.259$, $L_{\mathrm s}=1000$, and $\kappa^2\ge14.4$, i.e.\ $\kappa\ge3.79$ and $K\ge16.3$. A constant-$R_s$ varactor has $\kappa=\Delta X/R_s$, so $\Delta X/R_s\ge3.79$ at $f_0$.
\emph{Continuous tuning.} If the passband edge is swept across $f_0$, the transmission at $f_0$ follows a path from $L_{\mathrm p}$ to $L_{\mathrm s}$, and the attenuator bound of Note~S12 applies at $f_0$ with the element budget at $f_0$. The condition must hold at every frequency of the tuning range; the highest frequency is usually decisive.
\emph{Switched filter banks} are SP$n$T switches followed by fixed filters; Note~S11 applies to the switches.
Selectivity and bandwidth at a fixed state are governed by resonator $Q$ and by Bode--Fano theory \cite{Fano1950a,Fano1950b}; they are separate constraints.

\section{Reconfigurable Antennas and Metasurfaces}\label{sec:S14}
\subsection{Antenna as a contractive response}
Expand the radiated field in spherical vector modes normalized so that $|y_n|^2$ is the power in mode $n$ per unit incident power at the feed. Truncating to the modes that carry the radiated power gives a finite vector $y=Fu$; the reflected power and absorbed power are the remainder, so $\|y\|^2\le1$ is the total efficiency (radiation efficiency times mismatch efficiency). The normalized pattern is $\psi=y/\|y\|$ and the far-field correlation magnitude of two patterns is $c=|\langle\psi_1,\psi_2\rangle|$, the inner product of their normalized far fields over the sphere. The tuning elements on the antenna form the tuner; the radiating structure is the network.

\subsection{Pattern switching}
Note~S11 applies verbatim: switching between patterns with correlation $c$ at total efficiency $\eta$ requires settings $\delta_{\eta,c}$ apart. For $c=0.3$, $\eta=0.8$: $\delta=2\acosh(3.8)=4.02$.

\subsection{Continuous steering}
Write $y=\sqrt\eta\,\psi$ with $\|\psi\|=1$, so that $\dd y=(\dd\eta/2\sqrt\eta)\,\psi+\sqrt\eta\,\dd\psi$. Because $\|\psi\|=1$, $\langle\psi,\dd\psi\rangle$ is imaginary; write it $ja$ with real $a$. It measures how fast the overall phase of the pattern turns. The rest of $\dd\psi$ is orthogonal to $\psi$ and describes a change of shape; its squared length $\dd s_{\mathrm{FS}}^2=\|\dd\psi\|^2-a^2$ is the Fubini--Study element \cite{Bengtsson2006}. Hence $\langle y,\dd y\rangle=\dd\eta/2+j\eta a$ and $\|\dd y\|^2=\dd\eta^2/(4\eta)+\eta(\dd s_{\mathrm{FS}}^2+a^2)$, the cross term $\Re[\dd\eta\,ja]$ being zero. The vector metric $4[(1-\|y\|^2)\|\dd y\|^2+|\langle y,\dd y\rangle|^2]/(1-\|y\|^2)^2$ becomes
\begin{equation}
\dd\Lambda_y^2=\frac{\dd\eta^2}{\eta(1-\eta)^2}+\frac{4\eta\,\dd s_{\mathrm{FS}}^2}{1-\eta}+\frac{4\eta a^2}{(1-\eta)^2}:
\end{equation}
power change, pattern change and coherent phase change. Dropping the first and last terms with $\eta\ge\eta_{\min}$ gives $\Lambda\ge2\sqrt{\eta_{\min}/(1-\eta_{\min})}L_{\mathrm{FS}}$, and $L_{\mathrm{FS}}\ge\arccos c$, the Fubini--Study distance between the end patterns. For $c=0.3$, $\eta=0.8$: $\Lambda\ge5.06$.

\subsection{Efficient endpoints versus efficient throughout}
For orthogonal end patterns $e_0,e_1$ at efficiency $\eta$, the chord $y(q)=\sqrt\eta[(1-q)e_0+qe_1]$ has squared norm $\eta[(1-q)^2+q^2]$, dips to $\eta/2$, and has length $L_{\mathrm g}=2\acosh[1/(1-\eta)]$, equal to the distance between the ends by \eqref{eq:vecdist}; it is therefore a shortest path. Holding $\|y\|^2\ge\eta$ throughout requires $L\ge\pi\sqrt{\eta/(1-\eta)}$, attained by the constant-power rotation $\sqrt\eta(\cos\theta\,e_0+\sin\theta\,e_1)$. At $\eta=0.9$ the two lengths are $5.986$ and $9.425$. As $\eta\to1$, $L_{\mathrm g}=2\ln[2/(1-\eta)]+o(1)$ while the constant-power length $\pi\sqrt{\eta/(1-\eta)}\sim\pi/\sqrt{1-\eta}$.

\subsection{Metasurface cells}
A cell under periodic boundary conditions with one propagating Floquet mode per polarization is a multiport whose Floquet ports are the external ports; Notes~S7, S9 and S10 apply with the Floquet reflection as the selected response. Mutual coupling and fixed losses belong to the network \cite{Gradoni2021}.

\section{Multiport Networks and Arrays}\label{sec:S15}
\subsection{Uniformly attenuated transformations}
For $S=e^{-\ell}U$ with $U$ unitary, both defects equal $\sqrt{1-e^{-2\ell}}\,I$ and $\dd S=e^{-\ell}U(-\dd\ell\,I+W)$ with skew-Hermitian $W=U^\dagger\dd U$, eigenvalues $j\theta_i$. Since $-\dd\ell I+W$ is normal, its norm is $\max_i\sqrt{\dd\ell^2+\theta_i^2}=\sqrt{\dd\ell^2+\|W\|_2^2}$, so
\begin{equation}
\dd\Lambda_S=\frac{\sqrt{\dd\ell^2+\|U^\dagger\dd U\|_2^2}}{\sinh\ell},
\end{equation}
and the proofs of Note~S7 apply with $|\dd\phi|$ replaced by $\|U^\dagger\dd U\|_2$.

\subsection{Endpoint law}
For $S_i=\rho U_i$ with the same $\rho=e^{-\ell}$, evaluate \eqref{eq:mdist} in the eigenbasis of $U_0^\dagger U_1$ (eigenvalues $e^{j\theta_i}$). Both settings carry the same loss, so in that basis the matrix distance splits into independent scalar problems, one per eigenvalue, and the largest of them decides. Each eigen-direction gives the pseudo-hyperbolic value $q(\theta)=2\rho|\sin(\theta/2)|/\sqrt{(1-\rho^2)^2+4\rho^2\sin^2(\theta/2)}$, increasing in $|\theta|$, with $q/\sqrt{1-q^2}=|\sin(\theta/2)|/\sinh\ell$. Hence $d(S_0,S_1)=2\asinh[\sin(\Theta/2)/\sinh\ell]$ with $\Theta=\max_i|\theta_i|\in[0,\pi]$, where $\theta_i=\operatorname{Arg}\lambda_i(U_0^\dagger U_1)$ are the principal eigenphases.

\subsection{Rank test}
By \eqref{eq:fd}, $\rank[F(Y)-F(X)]\le\rank(Y-X)\le$ number of tuner ports. For a reciprocal $2\times2$ switch matrix with transmission block $T$, $S=\bigl[\begin{smallmatrix}0&T\\T^{\mathsf T}&0\end{smallmatrix}\bigr]$ and $\rank\Delta S=2\rank\Delta T$. Bar to in-phase cross: $\Delta T=I-\sigma_x$, rank~1, so $\rank\Delta S=2$. Bar to quadrature cross: $\Delta T=I-j\sigma_x$, $\det\Delta T=2$, rank~2, so $\rank\Delta S=4$: at least four tuner ports.

\subsection{Arrays}
An array with independent phase shifters has a block-diagonal tuner. Each element obeys Note~S7 for its own phase travel; the joint path length of a simultaneous sweep lies between the largest and the sum of the element lengths.

\section{State Counts and Cascades}\label{sec:S16}
\subsection{Packing and covering}
For a set $\mathcal S$ of settings let $\mathcal P_\delta(\mathcal S)$ be the largest number of points pairwise at least $\delta$ apart and $\mathcal N_\epsilon(\mathcal S)$ the least number of $\epsilon$-balls that cover it. For a $1$-Lipschitz map $F$: preimages of separated outputs are separated, so $\mathcal P_\delta(F(\mathcal S))\le\mathcal P_\delta(\mathcal S)$; images of covering balls lie in balls of the same radius, so $\mathcal N_\epsilon(F(\mathcal S))\le\mathcal N_\epsilon(\mathcal S)$.
\emph{One path.} States in path order with adjacent separation $\delta$ need $(M-1)\delta\le\Lambda$, so $M\le1+\lfloor\Lambda/\delta\rfloor$; a geodesic with equally spaced states attains it.
\emph{Independent elements.} With the maximum metric, the product of geodesic intervals is an $\ell_\infty$ box with packing number $\prod_k(1+\lfloor\Lambda_k/\delta\rfloor)$: a grid of step $\delta$ gives the lower bound, and assigning to each point the index tuple $\lfloor x_k/\delta\rfloor$ gives the upper bound, because two points with the same tuple are closer than $\delta$.
\emph{Meaning of $\delta$.} For reflection responses, by Note~S3, a source designed for one state and a load in another state $\delta$ away have mismatch efficiency $\sech^2(\delta/2)$. For patterns, vectors or codewords this number is a geometric measure of separation, not an operational efficiency: two orthogonal patterns at efficiency below one are a finite distance apart although their mutual projection is zero. A codebook should be scored with its own error metric, with $\delta$ used to count how many states a budget can hold apart.

\subsection{Uniform range contraction}
If $F$ is analytic between matrix balls and $\|F(X)\|_2\le r<1$ on the whole input ball, then $\dd\Lambda_F\le r\,\dd\Lambda_X$. In words, a fixed stage whose outputs all stay at least $1-r$ away from the lossless boundary shrinks every length by at least the factor $r$. Proof: for $r<\varrho<1$ write $F=\varrho G$ with $G$ into the open ball; Schwarz--Pick for the infinitesimal metric $\mu_{\mathrm{CK}}(X;\dd X)=\|D_{X^\dagger}^{-1}\dd X\,D_X^{-1}\|_2$, half the line element (5) of the main text, gives $\mu_{\mathrm{CK}}(G;\dd G)\le\mu_{\mathrm{CK}}(X;\dd X)$; the matrices $(I-\varrho^2YY^\dagger)^{-1/2}(I-YY^\dagger)^{1/2}$ and $(I-Y^\dagger Y)^{1/2}(I-\varrho^2Y^\dagger Y)^{-1/2}$ have eigenvalues $\sqrt{(1-\sigma^2)/(1-\varrho^2\sigma^2)}\le1$, so $\mu_{\mathrm{CK}}(\varrho Y;\varrho H)\le\varrho\,\mu_{\mathrm{CK}}(Y;H)$; let $\varrho\downarrow r$. Composition multiplies the radii. The radius is a property of the whole response range, not the insertion loss at one setting. For $F=r\gamma$, the ratio on a circle of radius $a$ is $r(1-a^2)/(1-r^2a^2)$; with $a=0.72$, $r=0.58$ the lengths are $18.79$ and $6.36$ (Fig.~1a of the main text).

\section{\texorpdfstring{The $\Lambda$-Chart: Construction and Use}{The Lambda-Chart: Construction and Use}}\label{sec:S17}
\subsection{Purpose}
\begin{figure}[p]
\centering
\includegraphics[width=\textwidth]{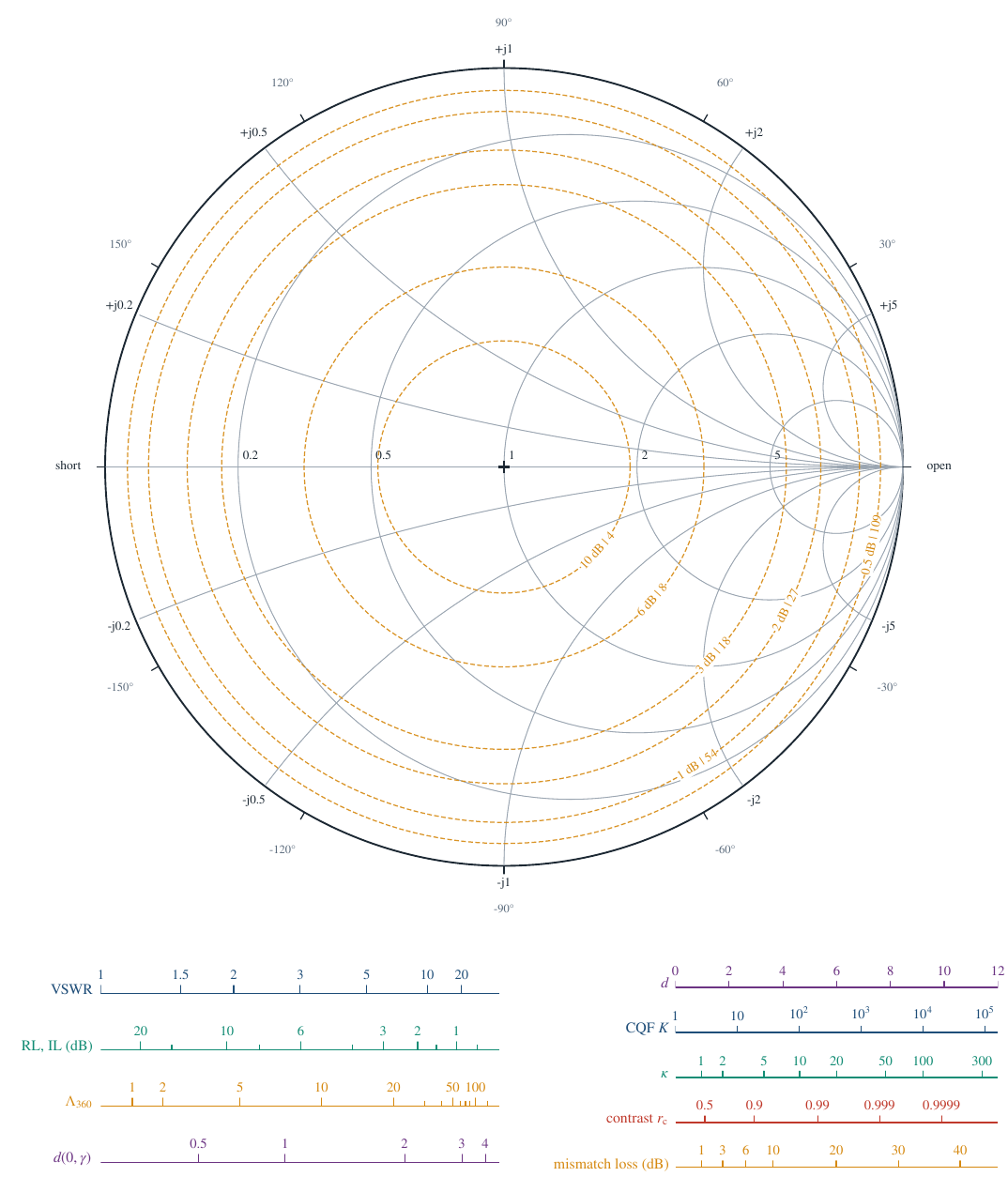}
\caption{Printable $\Lambda$-chart. Smith grid with normalized resistance and reactance values, short and open points and the angle of the reflection coefficient; price rings labeled with insertion loss and $\Lambda_{360}$; left rulers scaled by radius, with the same length as the chart radius, and right rulers scaled by the distance $d$ between two states. A radius is transferred to the left rulers with dividers; the distance $d$ read there is then carried to the right rulers.}
\label{fig:blank}
\end{figure}

The Smith chart lets an engineer move impedances through lossless networks by sliding along circles. The $\Lambda$-chart lets the engineer read, on the same disk, how much reconfiguration a tuning element can supply and how much a specification demands. It is a reflection-coefficient disk with the Smith grid and four layers, and two sets of rulers below it (Fig.~2 of the main text; a blank printable chart is Fig.~\ref{fig:blank}).

\subsection{Layers and their equations}
\begin{enumerate}
\item \emph{Smith grid.} Constant-resistance circles (horocycles) and constant-reactance arcs (geodesics), Note~S2. Impedances are plotted exactly as on a Smith chart, in any convenient reference.
\item \emph{Price rings.} Circles $|t|=\rho$ for selected insertion losses. Each ring carries two labels: $\IL=-20\log_{10}\rho$ and the full-turn price $\Lambda_{360}=4\pi\rho/(1-\rho^2)$ (Note~S7). The price of a phase travel $\Phi$ on the ring is $\Lambda_{360}\Phi/2\pi$.
\item \emph{Distance from the center.} The same circles read as distance from the center, $d=2\atanh\rho$, which is the logarithm of the VSWR; the left distance ruler gives it for any radius.
\item \emph{Contrast circles.} Around a chosen state $\gamma_1$, the circles \eqref{eq:hcircle} for $d=0.5,1,1.5,\dots$, drawn for the problem at hand and not printed on the blank sheet. The family shrinks toward the boundary, showing how a fixed budget gives less Euclidean motion near the lossless edge.
\item \emph{Budget ticks.} Along a plotted element trajectory, ticks every $\Delta\Lambda$ (e.g.\ $5$ units), placed by integrating $2|\dd\gamma|/(1-|\gamma|^2)$ or $|\dd Z|/\Re Z$. Two closed forms cover most elements: along a constant-resistance circle ($R_s$ fixed) a tick is a step $\Delta x=R_s\Delta\Lambda/Z_0$ of normalized reactance, and for a constant-$Q$ capacitor it is a capacitance ratio $e^{\Delta\Lambda/\sqrt{1+Q^2}}$. Near the rim the ticks crowd; for long trajectories take $\Lambda$ from Table~I of the main text and use the chart for the requirement side.
\end{enumerate}
Below the disk, two sets of rulers (Table~\ref{tab:rulers}) convert chart readings into the quantities used in the design rules. The left set is scaled by radius and is read with dividers, as on a Smith chart: VSWR, return or insertion loss, $\Lambda_{360}$ and the distance $d(0,\gamma)$ from the center. The right set is scaled by the distance $d$ between two states and gives $K$, $\kappa$, $r_{\mathrm c}$ and the mutual mismatch loss. No ruler belongs to a particular device; the floors of switches, bits, routers and filters follow from $d$ or $\kappa$ through the design rules (Table~II of the main text). To use it, read a contrast radius $r_{\mathrm c}$ on the left distance ruler, which gives $d=2\atanh r_{\mathrm c}$, and find the same $d$ on the right set. Both sets are linear: the left rulers in radius, from $0$ at the left end to $1$ at the right end, and the right rulers in $d$, from $0$ to $12$ on the printed sheet. Above $r_{\mathrm c}\approx0.95$ the radius is too close to the rim to read; compute $\kappa=|Z_1-Z_2|/\sqrt{R_1R_2}$ and $d=2\asinh(\kappa/2)$ from the impedances and enter $d$ directly.

\subsection{The re-centering operation}
Any lossless network moves the chart by an automorphism (Lemma~\ref{lem:lossless}). To read the contrast of a state $\gamma_2$ relative to $\gamma_1$, apply $\varphi(\gamma)=(\gamma-\gamma_1)/(1-\gamma_1^*\gamma)$: $\gamma_1$ goes to the center and $\gamma_2$ to radius $r_{\mathrm c}$. The map is not a rigid motion of the paper, so contrast circles drawn about one state do not serve another. In practice, compute $r_{\mathrm c}=|(Z_1-Z_2)/(Z_1+Z_2^*)|$ directly from the two impedances, or plot $\varphi(\gamma_2)$, and read the result on the rulers as above. Re-centering is free: it is what a lossless matching network does.

\subsection{Procedures}
\begin{figure}[p]
\centering
\includegraphics[width=0.92\textwidth]{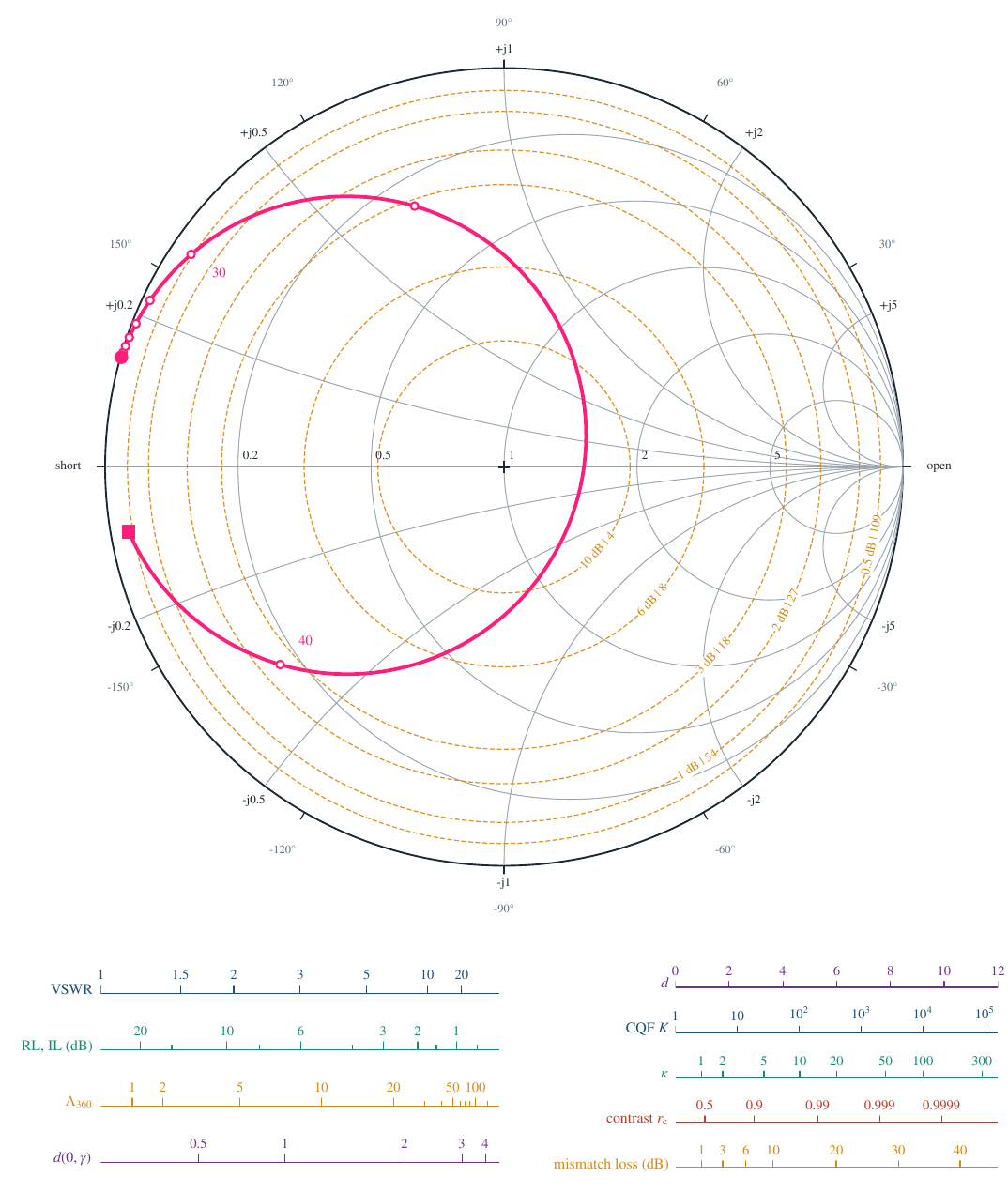}
\caption{Worked example on the printable chart, step~1: plot the element. Reflection of the single-varactor cell model of \cite{Abeywickrama2020} at $2.4$\,GHz ($Z_0=377\,\Omega$), from $C=0.47$\,pF (circle) to $C=2.35$\,pF (square), with a tick every $\Delta\Lambda=5$. The ticks crowd where the trajectory runs close to the rim: there the varactor is nearly lossless, and a small move costs a large length. The numbers give the length accumulated from the start: nine full intervals and a short remainder, $\Lambda=45.15$ by integration. On the angle scale the trajectory starts at $164^\circ$, passes through $0^\circ$ and ends at $-170^\circ$, so its phase travel is $164^\circ+170^\circ\approx334^\circ$.}
\label{fig:step1}
\end{figure}

\begin{figure}[p]
\centering
\includegraphics[width=0.92\textwidth]{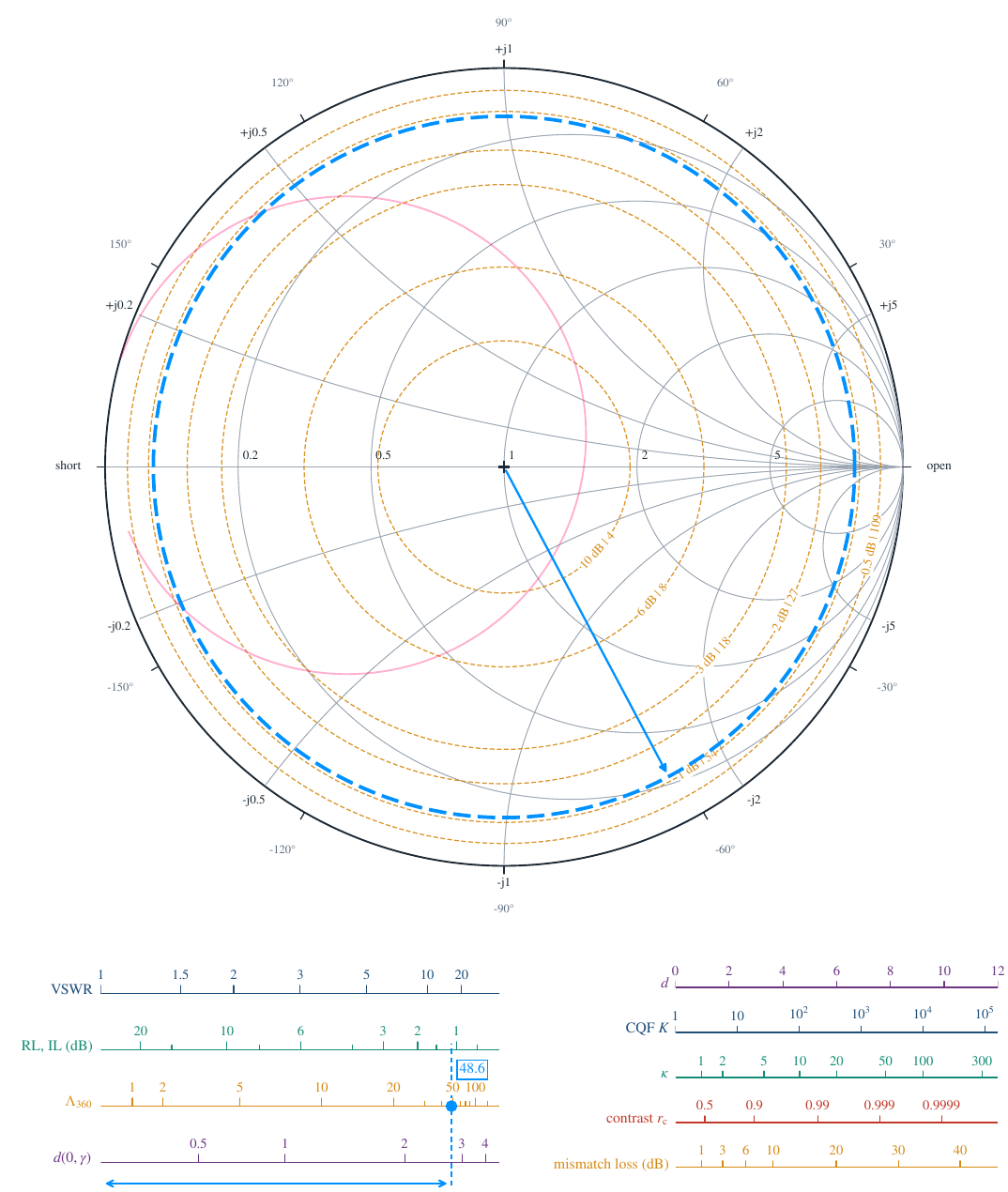}
\caption{Step~2: draw the ceiling. To travel $334^\circ$ on a ring of constant loss within the budget, a full turn on that ring may cost at most $\Lambda_{360}=360^\circ\times\Lambda/\Phi=360\times45.15/334.2=48.6$. Find $48.6$ on the $\Lambda_{360}$ ruler, set the dividers from the ruler's left end to that point (arrow), and swing them onto the chart from the center: the dashed circle of radius $0.879$. No passive cell built with this varactor can keep its amplitude above this circle over the whole $334^\circ$. The same dashed line crosses the loss ruler at $1.12$\,dB, the smallest peak loss any such cell can have. The faint curve is the cell of step~1.}
\label{fig:step2}
\end{figure}

\begin{figure}[p]
\centering
\includegraphics[width=0.92\textwidth]{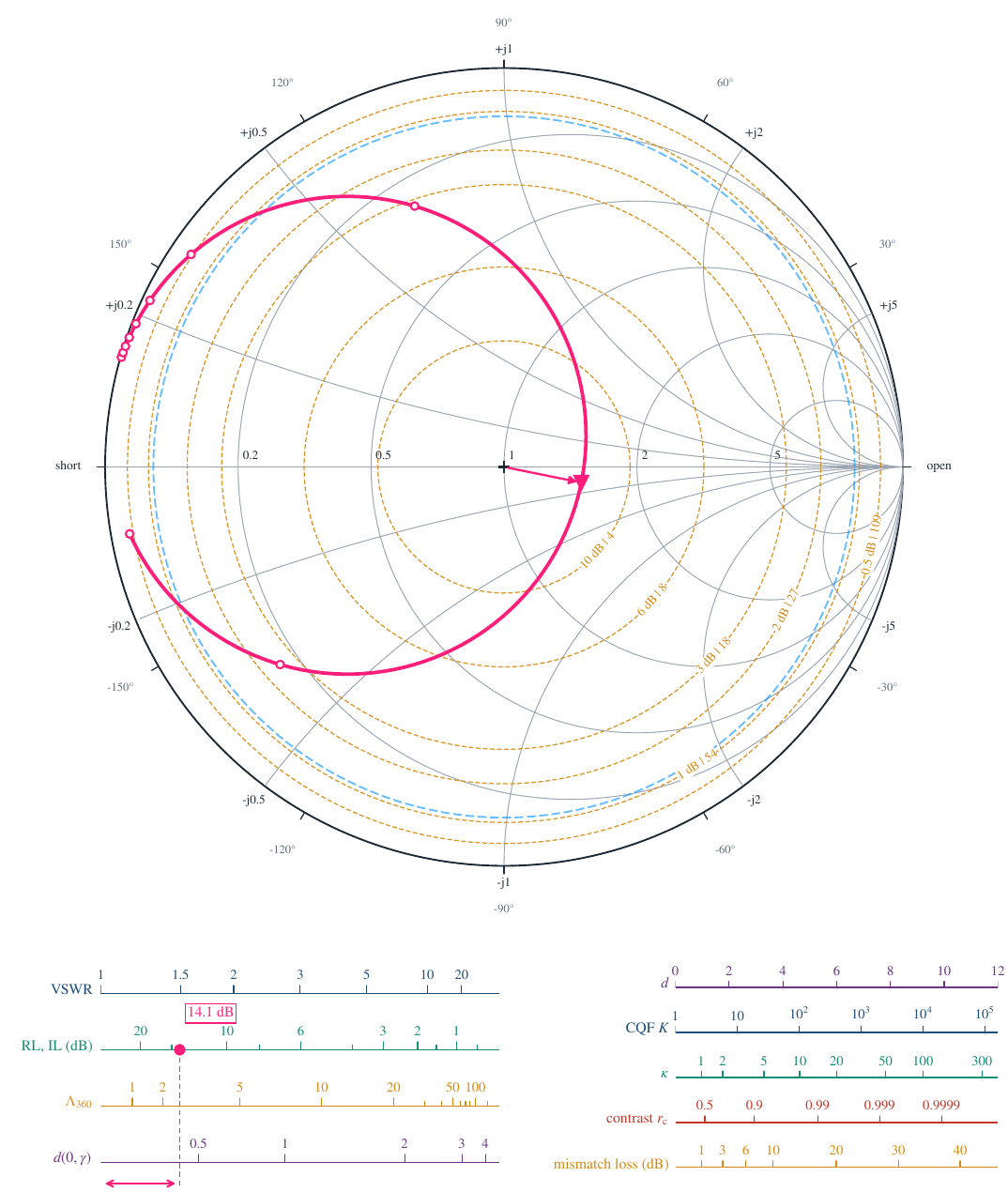}
\caption{Step~3: read the worst amplitude of the given cell. The point of the trajectory nearest the center (triangle) lies at radius $0.198$. Transferred with the dividers to the loss ruler, it reads $14.1$\,dB, far below the ceiling of step~2 (faint dashed circle).}
\label{fig:step3}
\end{figure}

\begin{figure}[p]
\centering
\includegraphics[width=0.92\textwidth]{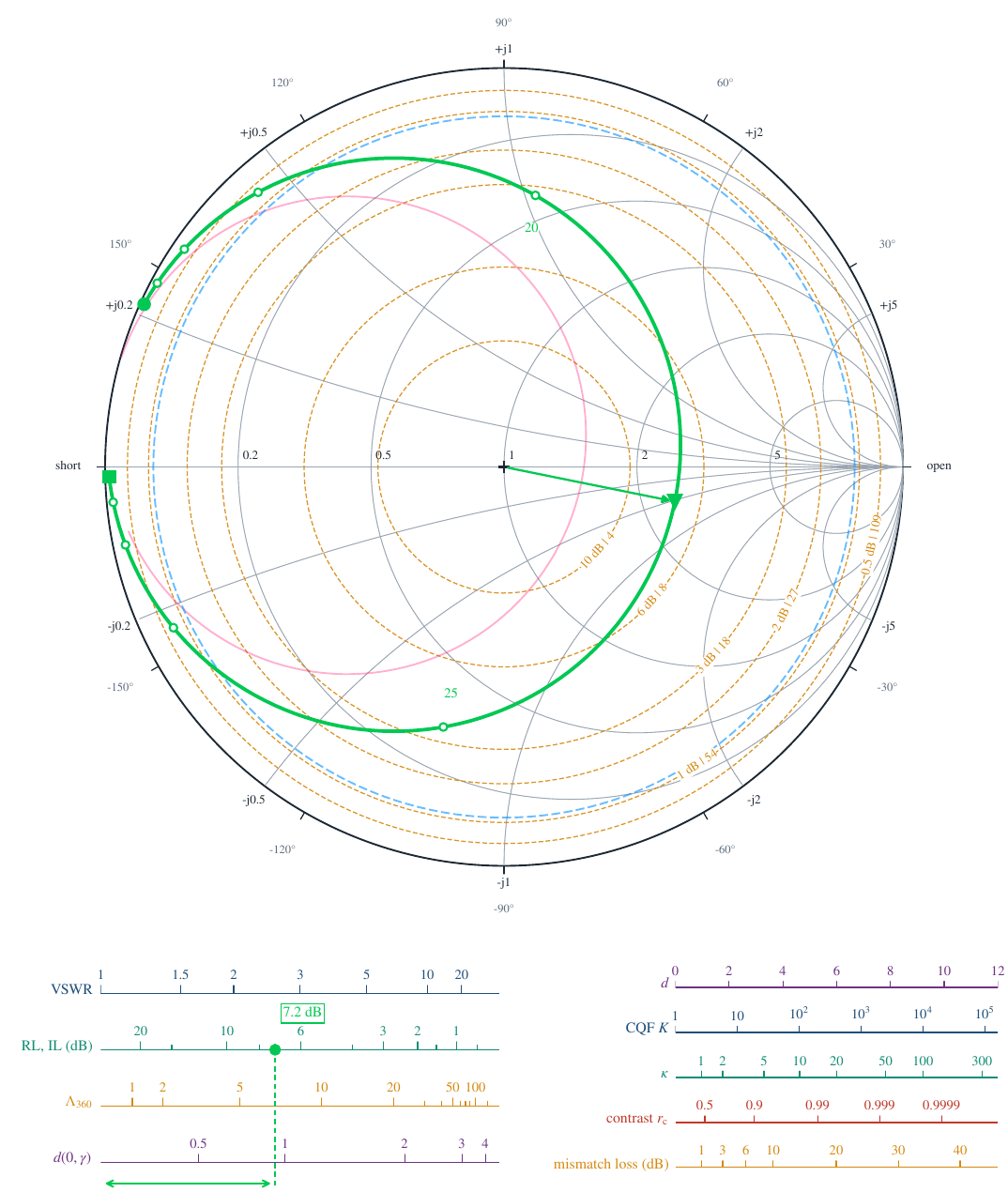}
\caption{Step~4: place the same varactor better. A lossless cell can carry the varactor's horocycle onto any other horocycle segment of the same length (Note~S9). The symmetric segment (numbers give the length from its start), with the same nine intervals and a short remainder and the same $334^\circ$ of phase, comes no closer to the center than $0.437$ (triangle), which reads $7.2$\,dB on the loss ruler. The worst-case loss drops from $14.1$ to $7.2$\,dB with the same varactor; the cell that does it is given in E2. The faint curve is the original cell and the faint dashed circle the ceiling.}
\label{fig:step4}
\end{figure}

\emph{A. Continuous phase specification.}
(1) Plot the element trajectory and place ticks every $\Delta\Lambda$; count ticks to obtain $\Lambda$.
(2) Find the price ring of the allowed peak loss; read $\Lambda_{360}$.
(3) Required budget $=\Lambda_{360}\Phi/2\pi$. If it exceeds an upper bound on $\Lambda$, the specification is excluded for this element model in every passive circuit. Otherwise it is not excluded; passing a necessary condition does not yet give a circuit.
(4) Utilization $=$ required$/\Lambda$.
\emph{B. Two-state specification.}
(1) Compute $r_{\mathrm c}$ or $d$ from the two element impedances, or plot both states and re-center on one.
(2) On the right rulers read $d$, $K$ and $\kappa$, then apply the rule for the device at hand (Table~II of the main text): for example $|t_{\mathrm{on}}|\le r_{\mathrm c}$ for an SPST with ideal isolation, $|t|\le(\sqrt K-1)/(\sqrt K+1)$ for a $180^\circ$ bit, $1-\eta\ge2\sqrt K/(K+1)$ for an SPDT.
\emph{C. Single-element cell design (horocycle).} Figs.~\ref{fig:step1}--\ref{fig:step4} carry out this procedure on the printable chart, one step per figure, for the cell of E2.
(1) Plot the cell and tick it (Fig.~\ref{fig:step1}). The varactor's Smith-chart circle is a horocycle, and every lossless cell maps it to a horocycle with the same ticks; they give $\Lambda$ and $\Phi$.
(2) Draw the ceiling, the price ring of price $2\pi\Lambda/\Phi$ (Fig.~\ref{fig:step2}). It bounds the minimum amplitude that any passive cell can keep over the required phase travel; individual states of a cell may lie outside it.
(3) Read the worst amplitude of the given cell (Fig.~\ref{fig:step3}).
(4) Place the segment that wraps the origin symmetrically with the same $\Lambda$ and $\Phi$, using \eqref{eq:horolen} (Fig.~\ref{fig:step4}; Fig.~3(b) of the main text plots the solution). Its minimum amplitude is attained by an explicit lossless cell.
\emph{D. Impedance tuner coverage.}
(1) Draw the VSWR-$S$ circle (radius $(S-1)/(S+1)$, hyperbolic radius $\ln S$).
(2) Draw one tolerance disk of mismatch loss ML (hyperbolic radius $\varepsilon=2\acosh 10^{\mathrm{ML}/20}$).
(3) The number of switched states needed is at least the area ratio $(S-1)^2/[4S(10^{\mathrm{ML}/10}-1)]$. For continuous tuners use the exact-coverage or finite-tolerance form of Note~S12.
\emph{E. Frequency ceiling.} Budget ticks scale with $1/f$ for varactors: at frequency $f'$ each tick of the $f$-chart is worth $f/f'$ ticks. The highest frequency at which the tick count still reaches the required price is the ceiling.
\emph{F. Variable attenuator.}
(1) Mark the rings of the smallest and largest loss, $\ell_1$ and $\ell_2$; for losses not printed, transfer their radii $e^{-\ell}$ from the loss ruler.
(2) Read both radii on the left distance ruler. The shortest path between the two rings is radial, so the required budget is the difference of the two readings, $\ln[\tanh(\ell_2/2)/\tanh(\ell_1/2)]$ with $\ell$ in nepers (Note~S12).
(3) Compare with the element budget, as in Procedure~A.
\emph{G. Reach of an impedance tuner.}
(1) Plot the impedance the network presents at one setting, $Z_A$.
(2) Draw the circle of hyperbolic radius $\Lambda$ about it: with $\gamma_A$ the reflection of $Z_A$, its Euclidean center $c_D$ and radius $R_D$ follow from \eqref{eq:hcircle} with $\gamma_1=\gamma_A$ and $D=\Lambda$. The center lies between $\gamma_A$ and the chart center, not on $\gamma_A$. Every impedance the tuner can present lies inside; a load whose conjugate $Z_L^*$ lies outside cannot be matched by any passive network built around this element (Note~S12).
\emph{H. Budget of any tuning element.}
(1) Plot the element's impedance over its control range, from a datasheet, a model or measurement, in any convenient reference impedance; the budget does not depend on it.
(2) Place ticks by integrating $|\dd Z|/\Re Z$, or use the closed forms of Note~S6. A constant-resistance element runs along a constant-resistance circle of the grid, a constant-$Q$ element along a circular arc through the short and open points, and a switch is a pair of points, read with Procedure~B.
\emph{What the chart does not show.} It carries one complex response at one frequency. Matrix and vector responses, such as switch matrices, beam-forming networks and radiation patterns, are handled with the formulas of Notes~S14 and~S15; any single coefficient of them can still be plotted and read on the chart.

\section{Worked Examples from Published Devices}\label{sec:S18}
Each example states the type of its data. \emph{Measured}: a number reported from hardware. \emph{Model}: a circuit model published for the device. \emph{Simulated}: a full-wave or circuit simulation reported by the authors. \emph{Parametric}: an illustrative value with no specific device. A bound computed from a model holds for the device to the extent that the model describes its tuning element. One case is safe. If the model leaves out a fixed lossy part, such as a bias-independent series resistance, that part belongs to the fixed network, and absorbing it into the element can only lower the budget (Note~S6.4); the floors computed without it are then lower than the true ones and remain valid, only looser. Bias-dependent loss is different. It is part of the tuning motion, and leaving it out can shrink the modeled budget instead of enlarging it: an element $Z=R_0+r(q)+jX_0$ tunes only through $r(q)$, and a model without $r(q)$ has no motion at all. Such a model gives no guaranteed direction.
\subsection{\texorpdfstring{E1. Analog reflection-type phase shifter \cite{Burdin2015}}{E1. Analog reflection-type phase shifter}}
\emph{Data.} Measured phase and loss; published diode model. \emph{Device.} Reflection-type phase shifter with two $\pi$-network arms of three jointly biased MA46H071 varactors; measured relative phase shift $385^\circ$ at $2$\,GHz with insertion loss $0.98\pm0.58$\,dB, i.e.\ at most $1.56$\,dB, and a reported figure of merit of $246^\circ$/dB; bias $0.5$--$20$\,V. Published series-$RC$ model: $R_s=1\,\Omega$, $C=0.45$--$2.72$\,pF.
\emph{Budget.} $Q_{\max}=1/(2\pi\cdot2\times10^9\cdot1\cdot0.45\times10^{-12})=176.8$, $n=6.04$, $\Lambda\le176.8(1-1/6.04)=147.6$. The six diodes are identical and jointly driven, so they count once (Note~S6). The model range includes zero bias, so $147.6$ is an upper envelope for the measured sweep.
\emph{Requirement.} $\Phi=385\pi/180=6.720$ rad, $\ell_{\max}=1.56\ln10/20=0.1796$, $\sinh\ell_{\max}=0.1806$, $\Lambda\ge6.720/0.1806=37.2$.
\emph{Chart reading} (Procedure A). The price ring at $1.56$\,dB has $\rho=0.8356$ and $\Lambda_{360}=34.8$; $385^\circ$ costs $34.8\times385/360=37.2$, i.e.\ about $3.7$ ticks of $10$ units, against $14.8$ ticks of the varactor; near the rim these ticks crowd, so the varactor's length is taken from Table~I of the main text.
\emph{Findings.} Utilization at least $37.2/147.6=25\%$. FoM: reported $246^\circ$/dB ($385/1.56=246.8$), bound $6.596\cdot147.6\cdot\sinh(0.1796)/0.1796=979^\circ$/dB. Element count: $385^\circ>360^\circ$ needs $N\ge2$ (the device has six). Ceiling: $f\le353.7\,\mathrm{GHz}\cdot(1-1/6.04)\cdot0.1806/6.720=7.9$\,GHz. Component tolerance at $2$\,GHz: $\Lambda\propto1/R_s$, so $R_s\le147.6/37.2=3.97\,\Omega$. Peak-loss floor for $385^\circ$ with this diode: $8.686\asinh(6.720/147.6)=0.40$\,dB.

\subsection{\texorpdfstring{E2. Single-varactor cell model \cite{Abeywickrama2020,Zhu2013}}{E2. Single-varactor cell model}}
\emph{Data.} Published circuit model; the hardware it is based on was measured, but the numbers below are model results. \emph{Device model.} Cell impedance
\begin{equation}
Z_{\mathrm{cell}}=\frac{j\omega L_1\bigl(j\omega L_2+\frac1{j\omega C}+R\bigr)}{j\omega L_1+j\omega L_2+\frac1{j\omega C}+R},\qquad\Gamma=\frac{Z_{\mathrm{cell}}-Z_0}{Z_{\mathrm{cell}}+Z_0},
\end{equation}
$L_1=2.5$\,nH, $L_2=0.7$\,nH, $R=2.5\,\Omega$, $C=0.47$--$2.35$\,pF, $Z_0=377\,\Omega$, $f=2.4$\,GHz. These are the circuit-model values that \cite{Abeywickrama2020} uses for a varactor cell after the metasurface of \cite{Zhu2013}. The amplitude model fitted in \cite{Abeywickrama2020} has $\beta_{\min}=0.2$, $\phi=0.43\pi$ and $\alpha=1.6$.
\emph{Budget.} $Q_{\max}=1/(2\pi\cdot2.4\times10^9\cdot2.5\cdot0.47\times10^{-12})=56.44$, $n=5$, $\Lambda=56.44\cdot0.8=45.15$.
\emph{Check of losslessness.} The fixed structure ($L_1$, $L_2$, free-space port) is lossless, so the cell trajectory must have exactly the element length. Numerical integration of the cell trajectory gives $45.150$ (Note~S21). The cell spans $334.2^\circ$ with minimum amplitude $0.198$, matching the amplitude model of \cite{Abeywickrama2020} ($\beta_{\min}=0.2$).
\emph{Horocycle reading} (Procedure C, worked on the chart in Figs.~\ref{fig:step1}--\ref{fig:step4}). The trajectory is a horocycle segment with $2s-1=0.198$, $s=0.60$. Solving \eqref{eq:horolen} for a symmetric segment with the same length and phase travel gives $s=0.7185$, $\vartheta_1=17.9^\circ$, and minimum amplitude $0.437$; the end amplitude is $0.990$. The same varactor, placed by a different lossless cell, more than doubles the worst-case amplitude.
\emph{Explicit improved cell.} The automorphism that carries the varactor's own horocycle (reference $377\,\Omega$, tangent at $\gamma=1$) onto the symmetric segment, with its tangent point kept at $1$, is $\varphi(\gamma)=e^{j\theta}(\gamma-a)/(1-a^*\gamma)$ with $a=-0.8753-j0.4141$ ($|a|=0.9683$, $\arg a=-154.68^\circ$) and $\theta=335.10^\circ$, using the $e^{j\omega t}$ convention, in which a line of electrical length $\beta l$ multiplies $\Gamma$ by $e^{-2j\beta l}$. By Lemma~\ref{lem:lossless} it is realized, from the varactor outward, by a $377\,\Omega$ line of electrical length $102.66^\circ$, an ideal transformer that divides the impedance by $(1+|a|)/(1-|a|)=62.05$ toward the Floquet port (turns ratio $7.877{:}1$), and a $377\,\Omega$ line of $89.79^\circ$ ending at the Floquet port. Fig.~2(b) of the main text and Fig.~\ref{fig:step4} draw the response with this last line set to $5.51^\circ$, which rotates it by $+168.56^\circ$ onto the orientation of the conventional cell; a line changes only the phase reference. At $C=0.47$\,pF the reflection is $0.990\angle{-12.88^\circ}$, at $C=0.783$\,pF it reaches its minimum, $0.437\angle180.0^\circ$, and at $C=2.35$\,pF it is $0.990\angle12.88^\circ$, the phase passing through $180^\circ$ on the way. This is an ideal single-frequency realization. Its computed reflection has minimum amplitude $0.4371$, phase travel $334.24^\circ$, end amplitudes $0.990$ and trajectory length $45.150$, equal to the varactor budget (Note~S21). The conventional cell of \cite{Abeywickrama2020} and this cell use the same element under the same lossless assumption; their worst-case amplitudes are $0.198$ and $0.437$, and Rule~1 caps any passive cell at $0.879$. The ideal transformer stands for any lossless matching network with the same input reflection; a practical cell would synthesize it with lumped or distributed elements. Their loss shortens the trajectory, and Rule~1 still caps any such cell at $0.879$.
\emph{Ceiling.} Rule~1 gives, for $334.24^\circ$ ($5.8336$ rad), $\ell\ge\asinh(5.8336/45.150)=0.1288$, i.e.\ no passive cell holds an amplitude above $0.879$ ($1.12$\,dB) throughout. Element count: one varactor gives less than $360^\circ$, in agreement with the ``almost $2\pi$'' range of the cell.

\subsection{\texorpdfstring{E3. One-bit varactor RIS \cite{Pei2021}}{E3. One-bit varactor RIS}}
\emph{Data.} Published circuit model and simulated design states; measured phase range. \emph{Device model.} Two Skyworks SMV2019-079LF varactors per cell at $5.8$\,GHz, with the circuit-model table of \cite{Pei2021}; series model $C=2.31$, $0.55$, $0.27$ and $0.24$\,pF and $R=4.51$, $3.66$, $2.65$ and $2.38\,\Omega$ at $0$, $7$, $16$ and $19$\,V, with $L=0.70$\,nH. The two design states, $7$\,V and $16$\,V, come from the simulated cell response; the measured $180^\circ$ pair lies near $3.8$\,V and $16$\,V on the same sweep, and the measured sweep from $0$ to $16$\,V spans about $250^\circ$ with a $6.5$\,dB gain fluctuation.
\emph{Contrast} (Procedure B). $\omega=3.644\times10^{10}$\,s$^{-1}$; $Z_7=3.66-j49.9\,\Omega$, $Z_{16}=2.65-j101.6\,\Omega$ (the series inductance is lossless and belongs to the network). $\kappa^2=[(1.01)^2+(51.7)^2]/(3.66\cdot2.65)$ $=276$, $\kappa=16.6$, $d=2\asinh(\kappa/2)=5.63$, $K=e^d=278$, $r_{\mathrm c}=0.993$. The radius is too close to the rim to read, so $d=5.63$ is entered on the right rulers directly, where it meets $K\approx280$ and $\kappa\approx17$, as in Fig.~2(c) of the main text; the $180^\circ$-bit rule then gives the floor below.
\emph{Findings.} A $180^\circ$ cell with equal amplitudes needs $\sinh\ell\ge2/16.6=0.120$, i.e.\ $\IL\ge1.04$\,dB in both states, for every cell geometry. \emph{Continuous sweep, rigorous bound.} The mean-loss floor needs an \emph{upper} bound on the path length; a sum of geodesic distances between exact samples is a lower bound on it (Note~S6.7), and an interpolated trajectory depends on the interpolation and is in general neither a lower nor an upper bound. Assume only that $C(V)$ and $R(V)$ are monotone between the tabulated biases, as for a reverse-biased varactor. On each interval $|\dd Z|\le|\dd X|+|\dd R|$ and $R\ge\min(R_a,R_b)$, so $\Lambda_{ab}\le|X_b-X_a|/\min(R_a,R_b)+|\ln(R_a/R_b)|$. With $X_0=11.88$, $X_7=49.89$, $X_{16}=101.63\,\Omega$: $\Lambda_{0\text{--}7}\le10.59$, $\Lambda_{7\text{--}16}\le19.85$, so $\Lambda\le30.4$. The measured $250^\circ$ ($4.363$ rad) then costs at least $8.686\asinh(4.363/30.4)=1.24$\,dB phase-weighted mean loss. \emph{Model estimate.} Linear interpolation of the model in the impedance plane gives $\Lambda\approx25.9$ and $1.46$\,dB; this is an estimate, not a bound.

\subsection{\texorpdfstring{E4. Two-bit PIN RIS \cite{Dai2020}}{E4. Two-bit PIN RIS}}
\emph{Data.} Published circuit model and simulated cell loss. \emph{Device model.} Skyworks SMP1340-040LF PIN diodes at $2.3$\,GHz with the circuit model of \cite{Dai2020}: on, $R=0.8\,\Omega$; off, $C=202$\,fF and $R=10\,\Omega$; series $L=780$\,pH in both states (a model value; the package datasheet lists $1.5$\,nH, which is lossless and changes no distance).
\emph{Contrast.} $X_{\mathrm{off}}=-1/(2\pi\cdot2.3\times10^9\cdot202\times10^{-15})=-342.6\,\Omega$; $\kappa^2=[(9.2)^2+342.6^2]/(0.8\cdot10)=1.47\times10^4$, $\kappa=121.2$, $K\approx\kappa^2$.
\emph{Findings.} The five diodes form two control groups: PIN1 and PIN2 switch complementarily, while PIN3--PIN5 switch together. Every pair of distinct programmed cell states differs in at least one diode, so the tuner distance is the single-diode distance under the common diode model. Pairs $180^\circ$ apart with equal loss in both states need $\IL\ge8.686\asinh(2/121.2)=0.14$\,dB, pairs $90^\circ$ apart $0.10$\,dB. These are balanced-pair benchmarks; states of unequal loss are constrained jointly by \eqref{eq:master}, so one state may lose less if the other loses more. The simulated state losses of $0.8$--$1.2$\,dB reported in \cite{Dai2020} are far above the benchmarks, but the difference for an individual state is not its own extra dissipation. The bound does not say where the excess arises: it may come from the substrate, metal and bias network, or from a coupling to the diode that is far from the optimum of Note~S10, which dissipates more in the diode than its floor requires. Separating the two needs a loss decomposition of the cell, for example by full-wave simulation with the diode resistance set to zero.

\subsection{\texorpdfstring{E5. Phase-change switch technology \cite{ElHinnawy2013}}{E5. Phase-change switch technology}}
\emph{Data.} Reported switch parameters. The GeTe inline phase-change switch of \cite{ElHinnawy2013} has a measured ON-state resistance of $1.2\,\Omega$, DC OFF-state resistance of $112$\,k$\Omega$, and OFF-state capacitance of $18.1$\,fF. Its conventional switching cutoff-frequency figure of merit is $f_c=1/(2\pi R_{\mathrm{on}}C_{\mathrm{off}})=7.3$\,THz. Applying the present theory requires the complex RF impedances of both states, including their frequency-dependent dissipation. The conventional cutoff-frequency figure of merit alone does not establish the hyperbolic state separation or the corresponding insertion-loss floors. Two readings of the same data show why. If the off state is a series $R_{\mathrm{off}}=R_{\mathrm{on}}$ with $C_{\mathrm{off}}$, then $\kappa=f_c/f$, which is $244$ at $30$\,GHz and $73$ at $100$\,GHz. If it is the DC resistance in parallel with $C_{\mathrm{off}}$, its series resistance is about $X_{\mathrm{off}}^2/R_{\mathrm{off}}$ and $\kappa\approx306$ at both frequencies, so the floors would change very little between 30 and 100\,GHz. The RF loss of the off state, which neither number captures, decides the answer and has to be measured.

\subsection{E6. Antenna tuner (parametric)}
\emph{Data.} Parametric. A switched-capacitor tuner must match handset antennas out to VSWR $10$ with mismatch loss at most $0.1$\,dB. Procedure D: $(S-1)^2/(4S)=2.025$, $10^{0.01}-1=0.02329$, $M\ge86.9$, so at least $87$ states (for instance a $7$-bit bank, provided the states are well distributed). A two-varactor continuous tuner that presents every load of the VSWR-$10$ disk exactly needs $\Lambda_1\Lambda_2\ge32.4$; within the $0.1$\,dB allowance the finite-tolerance form of Note~S12 applies and asks for less.

\subsection{E7. Tunable-dielectric phase shifter (parametric)}
\emph{Data.} Parametric, constant loss tangent. A liquid-crystal line with $\tau=0.24$ and $\tan\delta=0.0143$ has $\Lambda\approx19.2$ (Note~S6). A full turn at constant loss then needs $\ell\ge\asinh(2\pi/19.2)=0.322$, i.e.\ $2.79$\,dB, and the figure of merit cannot exceed $6.6\cdot19.2\approx127^\circ$/dB at low loss. A ferroelectric capacitor with $Q=50$, $n=2$ has $\Lambda=34.7$ and needs at least $1.57$\,dB peak loss for $360^\circ$.

\subsection{E8. True-time-delay line (parametric)}
\emph{Data.} Parametric. Note~S7: $f_{cv}=350$\,GHz, $n=6$, $1$\,dB, $f_H=10$\,GHz gives $\Delta\tau\le53.6$\,ps; doubling $f_H$ divides this upper bound on the delay change by four.

\section{Other Wave Platforms}\label{sec:S19}
The theorem uses only power-normalized amplitudes and a passive fixed structure.
\emph{Photonics.} Power-normalized Jones matrices and the power-normalized scattering matrices of passive optics, with their selected input-to-output blocks, are contractions. A chain matrix that relates forward and backward waves across a device is not, and is not used here. A two-mode rotation $S=\rho R(\theta)$, $0\le\theta\le\pi/2$, has $\|R^\dagger\dd R\|_2=|\dd\theta|$, so with $\rho\ge0.9$ it requires $\Lambda\ge(\pi/2)/\sinh(-\ln0.9)=14.88$; equal-attenuation endpoints $0.9I$ and $0.9R(\pi/2)$ are $2\asinh[\sin(\pi/4)/\sinh(-\ln0.9)]=5.20$ apart.
\emph{Acoustics.} With pressure $p$ and volume velocity $U$, the carried power is $|p|^2\Re Y/2$; normalizing amplitudes by $\sqrt{\Re Y}$ gives contractive scattering, and $\dd\Lambda=|\dd Z_a|/\Re Z_a$ for an acoustic port. A load $Z_a=0.2+jX$ with $X$ from $-1$ to $1$ has $\Lambda=10$; a $90^\circ$ phase excursion at $|t|\ge0.75$ needs $\Lambda\ge5.39$.
\emph{Passive quantum optics.} A passive bosonic network with vacuum loss inputs has $\hat a_{\mathrm{out}}=S\hat a_{\mathrm{in}}+L\hat e_{\mathrm{in}}$, $SS^\dagger+LL^\dagger=I$ \cite{GoughZhang2015}; a single photon in mode $u$ survives in the observed modes with probability $\eta=\|Su\|^2$. Routing it between orthogonal modes at $\eta=0.9$ requires settings $2\acosh10=5.99$ apart (Note~S11).

\section{Scope}\label{sec:S20}
\emph{Boundary responses.} Inverse defects require strict contractions. A response with a lossless external channel is handled by scaling the external output rows by $s<1$, applying the theorem, and letting $s\to1$; a finite limit is a valid bound, a divergent one gives no constraint.
\emph{Lossless elements.} An ideal reactance lies on the boundary and has infinite budget. A real element has a finite budget when it moves smoothly over a finite control range and its loss stays above zero in every state, so that its trajectory stays a finite distance from the boundary.
\emph{Broadband operators.} The proof uses only bounded operators, positivity and a bounded feedback inverse, so it applies to frequency-domain multipliers with a uniform gap from the boundary \cite{Staffans2012,WeissStaffans2013}; at each frequency the scalar results apply, and the worst frequency decides. Bode--Fano limits are complementary.
\emph{Modulated circuits.} If the control varies during propagation, frequencies couple and energy can be exchanged with the modulation source. The theorem applies to the complete frequency-coupled operator only when that operator is passive.
\emph{Nonlinearity and power.} The tuner is described by small-signal scattering. Large-signal varactor operation changes the trajectory and is outside the model.
\emph{Active circuits.} Amplifiers break passivity; the theorem applies to each passive section between active stages.

\section{Numerical Checks}\label{sec:S21}
Each check below was run in Python with NumPy/SciPy. 
\begin{enumerate}
\item Factorization (Theorem~1): $2000$ passive $5\times5$ networks with two external and three tuner ports, and $3\times3$ tuners, drawn at random with spectral norms uniform in $[0.2,0.999]$ for the network and $[0.2,0.98]$ for the tuner, and small random tuner steps; the largest local intrinsic gain was $0.34<1$.
\item Mismatch, VSWR and CQF identities, with reflection coefficients uniform in the disk $|\gamma|^2\le0.98$ and impedances with $R$ uniform in $[0.01,3]$ and $X$ normal with standard deviation $3$: agreement within rounding error in $1000$ random cases, and likewise for Kawakami's $r_{\mathrm c}=\tanh(d/2)$.
\item Varactor and constant-$Q$ budgets: numerical integration agrees with the closed forms.
\item Matrix impedance metric: agreement for random $3\times3$ impedances.
\item Attainment: the reflection-type construction of Note~S10 reproduces the $180^\circ$ bit ($|S_{21}|=0.78078$ at $\kappa=8$) and SPST ($|S_{21}|^2=1/(1+4/\kappa^2)$, off-state $S_{21}=0$) at the bounds with $S_{11}=0$.
\item Element count: random lossless networks of size $2+N$ terminated by $N=1,2,3$ varactors with normalized $R_s$ uniform in $[0.002,0.05]$ and a common capacitance swept geometrically over $2500$ points between random limits in $[0.05,0.5]$ and $[2,40]$ ($120$ networks each) gave at most $351^\circ$, $512^\circ$ and $597^\circ$ of phase travel.
\item Horocycle law: the E2 cell trajectory length equals the varactor length ($45.150$); the symmetric segment of Note~S9 has numerically integrated length $45.150$ and phase travel $334.2^\circ$.
\item Hyperbolic circles: points on \eqref{eq:hcircle} lie at distance $D$.
\item Chord length: numerical integration gives $5.98645$ at $\eta=0.9$, equal to $2\acosh10$.
\item Bias-dependent loss tangent: for $\tan\delta=0.01+0.02(C-1)^2$, $C\in[1,3]$, the directly integrated length is $58.2256$ and \eqref{eq:varQ} gives $58.2256$; the same integral without the derivative term, $\int\sqrt{1+Q(C)^2}\,\dd C/C$ with the local $Q(C)=1/\tan\delta(C)$, gives $58.04$.
\item Improved varactor cell (Note~S18, E2): the line--transformer--line embedding gives minimum amplitude $0.4371$, phase travel $334.24^\circ$ and length $45.150$; rebuilding it from the three elements reproduces the automorphism.
\item Rigorous Pei sweep bound: $\Lambda\le30.44$ and mean-loss floor $1.24$\,dB.
\end{enumerate}